\documentclass[aps,prd,nofootinbib,superscriptaddress]{revtex4-2}

\usepackage{amssymb}

\usepackage{mathrsfs}         
\usepackage{epsfig}
\usepackage{graphicx}
\usepackage{color}

\usepackage{amssymb}
\usepackage{fixltx2e}
\usepackage{bm}
\usepackage{amsmath,amsfonts,bm,dsfont}
\usepackage{verbatim}
\usepackage{mathrsfs}  %
\usepackage{definitions}
\usepackage{color} \definecolor{darkgreen}{rgb}{0,.5,0}
\usepackage{color}

\newcommand{\rv}[1]{{#1}}
\newcommand{\zz}[1]{{#1}}
\newcommand{\zg}[1]{{#1}}

\def\rA{{\rm A}}
\def\rR{{\rm R}}
\def\rK{{\rm K}}
\def\rM{{\rm M}}
\def\rT{{\rm T}}

\begin{document}
	\title{Discrete Wigner - Weyl calculus for the finite lattice}

	\author{M.A. Zubkov}
	\email{mikhailzu@ariel.ac.il}
	\affiliation{Ariel University, Ariel 40700, Israel}
	
	\date{\today}
	
	\begin{abstract}
		\centerline{\bf Abstract}
		We develop the approach of Felix Buot to construction of Wigner - Weyl calculus for the lattice models. We apply this approach to the tight - binding models with finite number of lattice cells. For simplicity we restrict ourselves to the case of rectangular lattice. We start from the original Buot definition of the symbol of operator. This definition is corrected in order to maintain self - consistency of the algebraic constructions. It appears, however, that the Buot symbol for simple operators does not have a regular limit when the lattice size tends to infinity. Therefore, using a more dense auxiliary lattice we modify the Buot symbol of operator in order to  build our new discrete Weyl symbol. The latter obeys several useful identities inherited from the continuum theory. Besides, the limit of infinitely large lattice becomes regular. We formulate Keldysh technique for the lattice models using the proposed Weyl symbols of operators. Within this technique the simple expression for the electric conductivity of a two dimensional non - equilibrium and non - homogeneous system is derived. This expression smoothly approaches the topological one in the limit of thermal equilibrium at small temperature and large system area.
	\end{abstract}
	
	\maketitle
	
	\tableofcontents

\section{Introduction}

The development of original Wigner-Weyl calculus began from the works of  H. Groenewold \cite{Groenewold1946} and J. Moyal \cite{Moyal1949}. This calculus replaces the conventional operator formulation of quantum mechanics by the formulation in terms of the Weyl symbols of operators. The Weyl symbol of operator is a function on phase space. This calculus was based on the ideas of H. Weyl \cite{Weyl1927} and E. Wigner \cite{Wigner1932}. Within the Wigner - Weyl calculus for quantum mechanics  Wigner distribution is used instead of the wave function, while Weyl symbols are used  instead of the operators of physical observables. Moyal product of the two functions defined in phase space replaces the product of two operators \cite{Ali2005,Berezin1972}. Applications of this calculus to several problems in quantum mechanics have been proposed \cite{Curtright2012,Zachos2005}. In the context of field theory certain modifications of Wigner-Weyl formalism were built. Such constructions were applied to the high energy physics theory and to the condensed matter physics \cite{Cohen1966,Agarwal1970,E.C.1963,Glauber1963,Husimi1940,Cahill1969,Buot2009}.
In particular, the notion Wigner distribution has been used in QCD \cite{Lorce2011,Elze1986}. In the framework of quantum kinetic theory the Wigner distribution has been used widely \cite{Hebenstreit2010,Calzetta1988}. The applications to the noncommutative field theories have been proposed as well \cite{Bastos2008,Dayi2002}.

Attempts to build lattice formulation of Wigner-Weyl calculus encountered certain difficulties. Some work in this direction has been undertaken already in the works by Schwinger \cite{Schwinger570}. Below we will discuss in more details the work by Buot \cite{Buot1974,Buot2009,Buot2013}. Here we would like to mention the physical applications discussed by Wooters \cite{WOOTTERS19871}, and Leonhardt \cite{Leonhardt1995}, as well as the mathematical constructions reported by  Kasperowitz \cite{KASPERKOVITZ199421}, and Ligab\'o \cite{Ligabo2016}. There were several other constructions of Wigner - Weyl calculus on the lattice - see, for example, \cite{BJORK2008469,GALETTI1988267,Cohendet_1988,doi,PhysRevA.53.3822,rivas1999weyl,mukunda2004wigner,chaturvedi2005wigner} and references therein. It is worth mentioning that the deformational quantization is related intimately to Wigner - Weyl calculus \cite{BAYEN197861,Kontsevich2003,Felder2000,Kupriyanov2008}.

The so - called approximate version of the lattice Wigner-Weyl calculus (more details see below) has been proposed in the works with the participation of the present author \cite{ZW2019}.
The application of this calculus is limited to the systems with weak inhomogeneity and slowly varying external fields. Using this formalism it is possible to express through the topological invariants the response of various nondissipative currents to external field strength \cite{Zubkov2017,Chernodub2017,Khaidukov2017,Zubkov2018a,Zubkov2016a,Zubkov2016b}. This formalism has also been applied to the investigation of scale magnetic effect \cite{Chernodub2016,Chernodub2017}.

Motivation of the present study originates from the idea to describe rigorously the topological properties of the non - homogeneous systems using Wigner - Weyl calculus. The methodology developed in the series of our previous publications \cite{ZW2019,FZ2020,ZZ2021,BFLZZ2021}, and in the present paper, may, in principle, be applied to the study of various non - dissipative transport phenomena. The corresponding conductivities are typically expressed as the topological invariants of various kinds. For definiteness we concentrate here at one of the most remarkable non - dissipative transport phenomena, which is the Quantum Hall Effect (QHE). The topological description of the QHE has been proposed originally for the uniform systems in the presence of constant external magnetic field. The conductivity of such an idealized system is expressed through the TKNN invariant \cite{Thouless1982}.  This invariant can also be used for the description of the   intrinsic anomalous quantum Hall effect (AQHE) in $2D$ topological insulators. This version of the QHE exists without external magnetic field. The TKNN topological invariant is expressed through the integral of Berry curvature over the occupied energy levels. If the one - particle Hamiltonian of the system is modified smoothly, then the TKNN invariant (i.e. the Hall conductivity) is not changed \cite{Avron1983,Fradkin1991,Hatsugai1997,Qi2008,Kaufmann:2015lga,Tong:2016kpv}.
Although expression for the QHE conductivity through the TKNN invariant is the most popular topological expression for the QHE, its applicability is limited to unphysically idealized systems, which are
uniform (except for the presence of external magnetic field) and non - interacting. This excludes, in particular, consideration of strong Coulomb interactions between electrons, the role of disorder, and boundary. The alternative topological description of the QHE may be given in terms of the Green functions. The development of the corresponding methodology has a long history. It began from the consideration of intrinsic AQHE in homogenous topological insulators without interactions  \cite{IshikawaMatsuyama1986,Volovik1988,Volovik2003a}. For the tight - binding model defined on the infinite lattice the corresponding expression is given by
$$
\sigma_H = \frac{\cal N}{2\pi},
$$
where
\be
	{\cal N}
	=  -\frac{ \epsilon_{ijk}}{  \,3!\,4\pi^2}\, \int d^3p \tr
	\[
	{G}(p ) \frac{\partial {G}^{-1}(p )}{\partial p_i}  \frac{\partial  {G}(p )}{\partial p_j}  \frac{\partial  {G}^{-1}(p )}{\partial p_k}
	\].
	\label{N-0}
	\ee
Here $G(p)$ is the two - point Green function depending on momentum. The physical meaning of $\cal N$ is that this is integer - valued topological quantity that provides quantization of Hall conductivity. 

The more simple topological invariant expressed through the two - point Green function is the one responsible for the stability of the Fermi surface \cite{Volovik2003a}
$
N_1= \tr \oint_C \frac{dp^l}{2\pi \ii}
G(p_0,p)\partial_l G^{-1}(p_0,p)
\label{N1}
$.
Here $\partial^\rho \equiv \frac\partial{\partial p_\rho}$, while $C$ is the contour that windes in momentum space around the Fermi surface. The more complicated topological invariant is responsible for the stability of  Fermi points   \cite{Matsuyama1987a,Volovik2003a}.
In fact, algebraically its expression is identical to the one of Eq. (\ref{N-0}). The similar constructions are also used in the other branches of condensed matter physics  \cite{HasanKane2010,Xiao-LiangQi2011,Volovik2011,Volovik2007,VolovikSemimetal}. In particular, such topological invariants protect gapless fermions existing along the boundaries of topological insulators \cite{Gurarie2011a,EssinGurarie2011} and in Dirac/Weyl semimetals \cite{Volovik2003a,VolovikSemimetal}, in the bulk. The similar constructions are also discussed for the  $^3$He-superfluid \cite{Volovik2016}. Various incarnations of these constructions may be found also in the high energy physics  \cite{NielsenNinomiya1981a,NielsenNinomiya1981b,So1985a,IshikawaMatsuyama1986,Kaplan1992a,Golterman1993,Volovik2003a,Hovrava2005,Creutz2008a,Kaplan2011}.

As it was mentioned above, the TKNN invariant has been obtained for the uniform system without interactions. The same refers also to expression of Eq. (\ref{N-0}). However, later it has been proven that Eq. (\ref{N-0}) remains valid in the presence of interactions when the two-point Green function is taken with the interaction corrections \cite{ColemanHill1985,Lee1986,ZZ2019}. Notice, that the original TKNN invariant cannot be extended in a similar way to the interacting systems as the matter of principle. However, the role of  interaction corrections to QHE conductivity was considered even before the mentioned above proof has been given   \cite{KuboHasegawa1959,Niu1985a,Altshuler0,Altshuler}.

Although Eq. (\ref{N-0}) solves the problem with interaction corrections, by construction it is defined for the homogeneous systems only. Direct extension of this expression to the non - homogeneous systems has been given in \cite{ZW2019}. Namely, for the tight - binding model of a two - dimensional system defined on infinite lattice
the Hall conductivity averaged over the system area is given by  $
\sigma_H = \frac{\cal N}{2\pi},
$ with
\be
{\cal N}
= - \frac{T \epsilon_{ijk}}{ |{\bf A}| \,3!\,4\pi^2}\, \int \D{{}^3x} \int_{\cM}  \D{{}^3p}
\, {\rm tr}\, {G}_{\cC}(x,p )\star \frac{\partial {Q}_{\cC}(x,p )}{\partial p_i} \star \frac{\partial  {G}_{\cC}(x,p )}{\partial p_j} \star \frac{\partial  {Q}_{\cC}(x,p )}{\partial p_k}.
\label{calM2d230I}
\ee
Here $T \to 0$ is temperature, $|{\bf A}| \to \infty $ is the area of the system, ${G}_{\cC}(x,p )$ is Wigner transformation of the two-point Green function $\hat G$. Correspondingly, ${Q}_{\cC}(x,p )$ is lattice Weyl symbol of operator $\hat{Q}$ ($\hat{Q}$ is inverse to the Green function itself). Symbol $\star$ means the Moyal product (see below Eq. (\ref{star0})). Notice that the Weyl symbol of an operator is defined for any real values of its arguments, and not only for the discrete coordinate lattice points $\vec{x}$. Eq. (\ref{calM2d230I}) has been derived originally for the non - interacting systems with weak inhomogeneity. This assumes, in particular, that the external magnetic fields remain not large (although are admitted to vary in space). Namely, the magnitude of magnetic field is much smaller than several thousands Tesla, while the wavelengths of external electromagnetic fields are much larger than $1$ Angstrom. Except for the artificial lattices, the bound on the magnetic field strength is satisfied for any real system.
 In \cite{ZZ2019_2,ZZ2021} it has been proven that in the presence of interactions the Hall conductivity is still given by the expression of Eq. (\ref{calM2d230I}), when the Green function is replaced by the complete interacting two-point Green's  function. The version of the Wigner-Weyl calculus for the lattice models with the Weyl symbol of operator $\hat A$ denoted above as $A_{\cC}$ was called approximate because some of the basic identities of continuum Wigner - Weyl calculus are satisfied by the given Weyl symbols only approximately \cite{Suleymanov2019}. The approximate nature of the formalism results in the requirement of the weakness of inhomogeneity. As a result Eq. (\ref{calM2d230I}) remains valid only approximately, and is to be modified for artificial lattices or in case of strong inhomogeneities. Notice, that these are the systems, where the so - called Hofstadter butterfly appears.

Extension of  Eq. (\ref{calM2d230I}) to the case of arbitrarily strong inhomogeneity has been given in \cite{FZ2020}. There the tight - binding model on the infinite rectangular lattice has been considered. The version of lattice Wigner - Weyl calculus has been proposed that was called "precise" because the corresponding Weyl symbol of an operator satisfies the basic identities of continuous Wigner - Weyl calculus precisely. Using the developed formalism it was derived that Eq. (\ref{calM2d230I}) is to be replaced by
\be
{\cal N}
= - \frac{ \epsilon_{ijk}}{ |{\bf A}| \,3!\,4\pi^2}\, \frac{|{\cal V}^{(2)}|}{2^{2} }\sum_{\overrightarrow x \in {\cal O}^\prime} \int_{\cM} {d^3p} \tr
\[
{G}_{W}(x,p )\star \frac{\partial {Q}_{W}(x,p )}{\partial p_i} \star \frac{\partial  {G}_{W}(x,p )}{\partial p_j} \star \frac{\partial  {Q}_{W}(x,p )}{\partial p_k}
\].
\label{calM2d230}
\ee
Here $x = (\tau, \overrightarrow{x})$, $\overrightarrow{x}$ is the point in space, $\tau$ is imaginary time that varies between $0$ and $1/T \to \infty$, but Wigner transformation of the Green function ${G}_{W}$ and Weyl symbol of its inverse ${Q}_{W}$ do not depend on $\tau$.  ${\cal V}^{(2)}$ is the area of the two-dimensional lattice cell while $|{\bf A}|\to \infty$ is the overall area of the system. By ${\cal O}^\prime$ we denote the refined lattice, in which the extra lattice sites are added with the half - integer coordinates (the coordinates of the original lattice are implied to be integer).

The silent feature of the construction proposed in \cite{FZ2020} is that Eq. (\ref{calM2d230}) contains the sum over the lattice of large but finite area  $|{\bf A}|$. Such a finite lattice approximates the infinite lattice. The limit of infinite  $|{\bf A}|\to \infty$ is to be taken at the end of calculation. At the same time the very definition of the Weyl symbol (entering Eq. (\ref{calM2d230})) as well as the derivation of its basic properties are valid for the infinite lattice with continuous compact Brillouin zone. Intuitively it is clear that  Eq. (\ref{calM2d230}) remains valid under these conditions. However, it is better to build the rigorous infrared regularization of the theory with the version of Wigner - Weyl calculus that obeys all necessary axioms inherited from the continuous theory, and that is reduced smoothly to the "precise" Wigner - Weyl calculus of \cite{FZ2020}. In the present paper we report the corresponding construction. Moreover, we extend the considerations of the Wigner - Weyl field theory to the kinetic domain, and build the version of Keldysh technique for the models defined on finite lattices, written on the language of Weyl symbols of operators instead of the conventional operator formalism. Here we follow the methodology developed earlier in \cite{Sugimoto,Sugimoto2006,Sugimoto2007,Sugimoto2008} for continuous models and extended in \cite{BFLZZ2021} to the lattice models with weak inhomogeneity using the mentioned above "approximate" Wigner - Weyl calculus.

The first attempt to build the version of Wigner  - Weyl calculus adopted for the models defined on finite lattices was performed in the works by Felix Buot  \cite{Buot1974,Buot2009,Buot2013}. He defined the Weyl symbol of an operator as a function on discrete phase space ${\cal O}\otimes {\cal M}$ composed of the original coordinate lattice ${\cal O}$ and discrete momentum space ${\cal M}$ with the same number of points. Unfortunately, the further constructions of \cite{Buot1974,Buot2009,Buot2013} contain certain algebraic inconsistencies. As a result, in particular, the so - called star property (see below Eq. (\ref{star0})) is  valid only approximately under the condition that external fields vary slowly (as well as in the mentioned above "approximate" Wigner - Weyl calculus of \cite{ZW2019}). Nevertheless, certain results remain valid precisely under the assumption that the proper lattice Weyl symbol is constructed and replaces the original definition by Buot. The very proposition to use lattice version of Wigner - Weyl calculus adopted for the finite lattice seems to us extremely important. Moreover, the ideas developed in the works of F.Buot are very interesting and fruitful. We pay tribute to the physical intuition of Felix Buot, to his constructions and physical results that remain correct in spite of certain technical mistakes. We consider our present study as a logical prolongation of his work.

We perform the correction of the original constructions by F.Buot in three steps. At the first step we build the symbol of operator that is called by us the Buot symbol of operator. This construction is maximally close to the original definition of F.Buot (see below Eq. (\ref{Buot0})). In principle, it is possible to build the self - contained version of lattice field theory completely in the language of the Buot symbol of operators. However, it appears that the limit of infinite lattice volume is not regular being written in terms of this type of the symbol of operator. The same refers to the attempt to approach continuum limit. More specifically, the Buot symbol of physically relevant operators (including the unit operator) is the fast oscillating function of both coordinates and momenta. As a result we proceed constructing the two subsequent modifications of the Buot symbol. The first modification results from the consideration of the model on the auxiliary more dense momentum lattice. The second modification is the mirror construction of the first modification based on the consideration of the auxiliary coordinate lattice that contains $2^D$ times more lattice points than the original lattice. The doubly modified Buot symbol still contains the fast oscillating factors depending on space coordinates and momenta. However, these factors are common for all considered operators. Our final construction appears when we simply omit these factors in the doubly modified Buot symbol. We call the resulting symbol of operator the Weyl symbol, and prove its basic properties. It appears, that this Weyl symbol obeys the basic axioms inherited from the continuous Wigner - Weyl calculus. Moreover, the limit of infinitely large lattice as well as the continuum limit are regular, i.e. the Weyl symbol as a function on phase space has the well - defined limit when the number of lattice points tends to infinity. As expected, this continuum limit is the Weyl symbol of "precise" Wigner - Weyl calculus proposed in \cite{FZ2020}. Written in terms of our new Weyl symbol the  Hall conductivity  of the two - dimensional system is given by  $
\sigma_H = \frac{\cal N}{2\pi},
$ with
\begin{eqnarray}
	{\cal N} &=& \frac{1}{3!\,}\epsilon^{\mu\nu\rho} \frac{1}{(2N)^{2D}} \int {d\Pi^3}  \sum_{p \in   {\cal M}^\prime, x\in {\cal O}^\prime}
	\tr\left(\partial_{\rv{\Pi^{\mu}}}\hat{Q}_W ^\rM  \star\hat{G}_W ^\rM \star \partial_{\rv{\Pi^{\nu}}}\hat{Q}_W ^\rM  \star\hat{G}_W ^\rM \star \partial_{\rv{\Pi^{\rho}}}\hat{Q}_W ^\rM \star \hat{G}_W ^\rM \right).
	\label{NEQ00}
\end{eqnarray}
Here $\hat{G}_W^M$ is the  Weyl symbol of  the Matsubara Green function while $\hat{Q}_W^M$ is Weyl symbol of its inverse {(it depends on the lattice size $N$ as on parameter, and the limit, when this parameter tends to infinity, is to be taken; correspondingly, $\hat{G}_W^M$ is to be calculated as inverse to this expression with respect to the Moyal product)}.
Momentum space is Euclidean one, its points are denoted by $\Pi_i$. In particular, $\Pi^3$ is the Matsubara frequency. The original coordinate space lattice $\cal O$ contains $N^2$ lattice sites, while the Brillouin zone $\cal M$ is discrete as well, and also contains $N^2$ points. By ${\cal O}^\prime$ we denote the twice more dense spatial lattice that contains $4 N^2$ points. Correspondingly, ${\cal M}^\prime$ is the twice more dense Brillouin zone containing $4 N^2$ points.   Eq. (\ref{NEQ00}) represents the rigorous infrared regularization of Eq. (\ref{calM2d230}).

The paper is organized as follows. In Sect. \ref{Statement} we start from the formulation of the main results. In Sect. \ref{BuotSymbol} we define the Buot symbol of operator and derive its basic properties. In Sect. \ref{ModifBuotSymbol} we consider the modification of the notion of Buot symbol and build the final Weyl symbol of operator defined on finite lattice. In Sect. \ref{Dynamics} we describe the dynamics of quantum kinetic theory written in terms of Weyl symbols, and derive the desired expression for the QHE conductivity. In Sect. \ref{Concl} we end with the conclusions.

\section{Statement of the main results}
\label{Statement}
\subsection{Definition of Weyl symbol and its properties}
\label{SectList}
\begin{figure}[h]
	\centering  %
	\includegraphics[width=14cm]{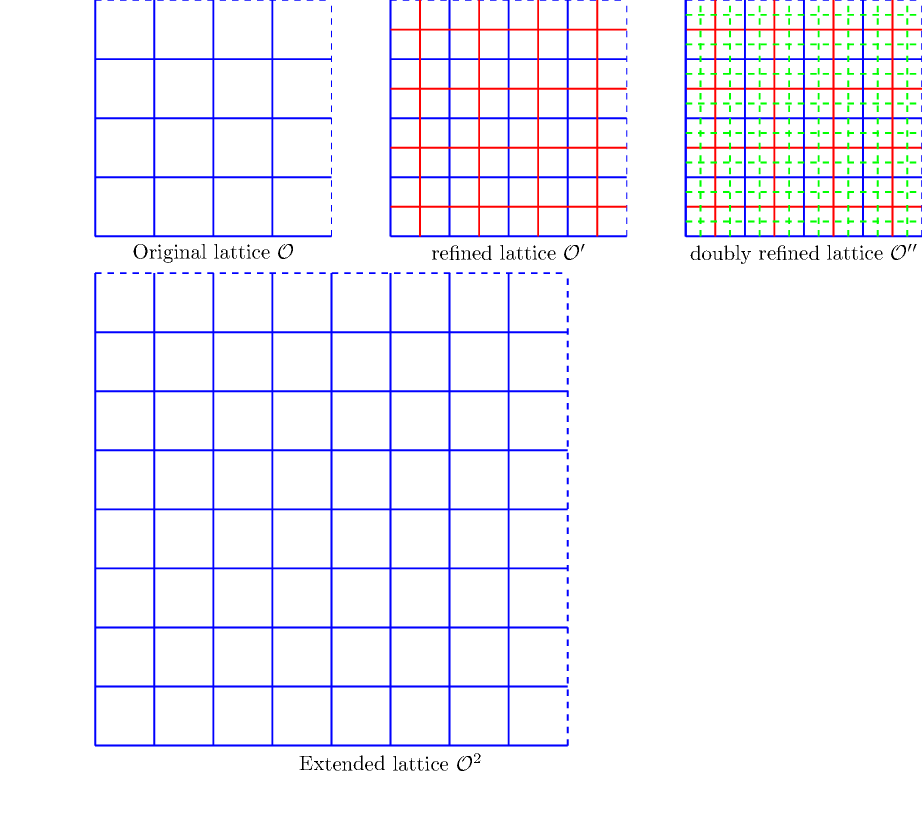} 
	\caption{In this figure lattices ${\cal O}, {\cal O}^\prime, {\cal O}^{\prime\prime}, {\cal O}^2$ are represented schematically. The  extended refined lattice ${\cal O}^{2\prime}$ and doubly refined extended lattice
		${\cal O}^{2\prime\prime}$ are defined in an obvious way combining both extention and the transition to the refined lattice.}  %
	\label{fig}   %
\end{figure}

We are considering the rectangular lattice
$$
{\cal O} = \{(m_1 ,...,m_D ) | m_i \in \{0,1,2,...,N-1\}\}
$$
with the number of points $N^D$ and periodic boundary conditions.
Momentum space is
$$
{\cal M} = \{(m_1 \frac{2\pi}{N},...,m_D \frac{2\pi}{N}) | m_i \in \{0,1,2,...,N-1\}\}
$$
Hilbert space ${\cal H}$ of one - particle states is spanned on ket vectors
$$
| q \rangle, q \in {\cal O}
$$
Another set of basis vectors is
$$
| p \rangle, p \in {\cal M}
$$
We assume here $| q +N \rangle = | q \rangle$ and $| p + 2\pi \rangle  = | p \rangle$.
Besides, we define the refined lattice
$$
{\cal O}^\prime = \{(m_1  ,...,m_D ) | m_i \in \{0,1/2,1,...,N-1/2\}\}
$$
with the number of points that is $2^D$ times larger. We also define the refined momentum space
$$
{\cal M}^\prime = \{(2\pi m_1/N,...,2 \pi m_D/N) | m_i \in \{0,1/2,1,...,N-1/2\}\}
$$
and extended refined spaces
$$
{\cal O}^{2\prime} = \{(m_1  ,...,m_D ) | m_i \in \{0,1/2,1,...,2N-1/2\}\}
$$
and
$$
{\cal M}^{2\prime} = \{(2\pi m_1/N,...,2 \pi m_D/N) | m_i \in \{0,1/2,1,...,2N-1/2\}\}.
$$
We illustrate these constructions by Fig. \ref{fig}, where lattices ${\cal O}, {\cal O}^\prime, {\cal O}^2$ are represented schematically. Besides, we represent in this figure the douply refined lattice ${\cal O}^{\prime\prime}$, which will be defined and used in the following sections. Lattices ${\cal O}^{2\prime}$ and
${\cal O}^{2\prime\prime}$ will also be used. Their definition, obviously, contains application of both extention and the transition to the refined lattice. The similar constructions refer also to various modifications of momentum space $\cal M$.

For $q \in {\cal O}^{\prime}, p \in {\cal M}^\prime$ we define Weyl symbol as follows
\begin{eqnarray}
	A_{{ W}}(p,q)  & = &
	\sum_{n_i = 0,1;  u \in {\cal M}^{\prime} }  \,
	\langle p + \pi n/N - u|\hat{A}|p + \pi n/N + u \rangle \,  e^{-2 i u q}\nonumber\\&&\Pi_i \frac{1 + e^{i u^i    }}{2}\frac{1+e^{N  i (p^i+  \pi n_i /N+ u^i)}}{2}
	\label{Weyl}
\end{eqnarray}
or, equivalently,
\begin{eqnarray}
	A_{{ W}}(p,q)  & = &
	\sum_{n_i = 0,1;v \in  {\cal O}^{\prime} } e^{2 i p v} \,
	\langle q-v {-} n/2 |\hat{A} |q+v {-}n/2 \rangle \nonumber\\ && \Pi_i\frac{{1 + e^{2i v_i \pi/(N) }}}{2}\frac{1+e^{2\pi i (q_i-v_i{-} n_i/2)}}{2}.
	\label{Weyl2}
\end{eqnarray}
Using periodicity of these expressions ($p_i\to p_i+2\pi, q_i \to q_i + N$) they are naturally extended to $q \in {\cal O}^{2\prime}, p \in {\cal M}^{2\prime}$.
This definition may be extended further to any continuous values of $q$ and $p$ as follows
 \begin{eqnarray}
	A_{{W}}(p,q)  & = & \sum_{p_1 \in {\cal M}^{\prime}; q_1 \in {\cal O}^{\prime}; p_2 \in {\cal M}^{\prime}; q_2 \in {\cal O}^{\prime}}\frac{1}{(4 N^2)^D} {e^{2 i  p_2(q_1-q) + 2 i q_2(p-p_1)}}   A_{{ {W}}}(p_1,q_1) .		\label{Wcont}
\end{eqnarray}

The matrix elements of operator $\hat A$ are expressed through Weyl symbol of this operator as follows:
\begin{eqnarray}
	\langle q_1 |\hat{A}|q_2\rangle  &=&{\frac{1}{(2N)^D}} \sum_{p \in {\cal M}^{\prime}}A_W(p,(q_1+q_2)/2)e^{ i p (q_1 - q_2) }\Pi_i \frac{2}{1+e^{i(q_2^i-q_1^i)\pi/N}}
\end{eqnarray}
and
\begin{eqnarray}
	\langle p_1 |\hat{A}|p_2\rangle &=&
	{\frac{1}{(2N)^D}}\sum_{ q \in {\cal O}^{\prime}} e^{ i q (p_2-p_1)} 	 A_W((p_2+p_1)/2,q)\Pi_i \frac{2}{1+e^{i(p_2^i-p_1^i)/2}}.
\end{eqnarray}

We formulate the following properties of the Weyl symbol:

\begin{enumerate}
	\item
	
	Trace property.
	$$	
	{\rm Tr}\, \hat{A} = \frac{1}{(4N)^D} \sum_{p \in {\cal M}^\prime , q\in {\cal O}^\prime } A_W(p,q)
	$$	
	
	\item

	Second trace identity.
	
	$$	
	{\rm Tr} \hat{A}\hat{B} = \frac{1}{(4N)^D} \sum_{p \in {\cal M}^\prime , q\in {\cal O}^\prime} A_W(p,q) B_W(p,q)
	$$	
	
	\item

	
	{Periodicity: \begin{equation*}
		A_W(p,q) - A_W(p+\tfrac{\pi}{N}{\bf e}_k,\,q) = e^{iNp^k}\Big[A_W(p,\,q+\tfrac{N}{2}{\bf e}_k) - A_W(p+\tfrac{\pi}{N}{\bf e}_k,\,q+\tfrac{N}{2}{\bf e}_k)\Big]\,,
		\label{per1}
	\end{equation*}
	\begin{equation*}
		A_W(p,q) - A_W(p,\,q-\tfrac{1}{2}{\bf e}_k) = e^{2\pi i q^k}\Big[A_W(p+\pi{\bf e}_k,\,q) - A_W(p+\pi{\bf e}_k,\,q-\tfrac{1}{2}{\bf e}_k)\Big]\,.
		\label{per2}
	\end{equation*}
	All arguments here remain on the lattice ${\cal M}'\times{\cal O}'$: the shifts $\tfrac{\pi}{N}{\bf e}_k$ and $\tfrac{1}{2}{\bf e}_k$ are one lattice step of the $k$--th component of the momentum and of the coordinate, respectively, while $\tfrac{N}{2}{\bf e}_k$ and $\pi{\bf e}_k$ translate these components by half of the corresponding period.}

	\item Star property
	
	\begin{eqnarray}
		&&	(\hat{A}\hat{B})_W(p,q)\Big|_{p \in {\cal M}^\prime , q\in {\cal O}^\prime }
		=  A_W(p,q)  e^{\frac{i}{2}(\overleftarrow{\partial_q}\overrightarrow{\partial_p}-\overleftarrow{\partial_p}\overrightarrow{\partial_q})}  B_W(p,q)=	 A_W(p,q)  \star B_W(p,q) \label{star0}
	\end{eqnarray}

Here derivative $\overrightarrow{\partial}$ acts to the right while $\overleftarrow{\partial}$ acts to the left. 
	
	\item Weyl symbol of unity
	
	$$
	1_W(p,q)\Big|_{p \in {\cal M}^\prime , q\in {\cal O}^\prime }      = 1
	$$
	
	\item
	Weyl symbol of translation to one lattice spacing is given by
	\begin{eqnarray}
		T_j(p,q)
		& = & e^{i p_j} \Big(\frac{1+{e^{i  \pi /N} }}{2}  + e^{i N p_j}\frac{1-{e^{i  \pi /N} }}{2}\Big)
	\end{eqnarray}
	One can see that the limit $N\to \infty$ of this expression exists and gives
	\begin{eqnarray}
		T_j(p,q)
		& \to & e^{i p_j}
	\end{eqnarray}

\item

In the limit of infinitely large lattice the Weyl symbol defined here smoothly approaches the Weyl symbol of \cite{FZ2020} given by
$$
A_{\cal W}(p,q) = \int_{{\cal M}} d^D p_-   \,
\langle\langle p  - p_-|\hat{A}|p  + p_- \rangle\rangle \,  {e^{-2 i p_- q}}\Pi_i (1 + e^{i p^i_-    }).
$$
Here vectors $|p\rangle \rangle$ are normalized as
$$
\langle \langle p_1|p_2\rangle \rangle =  \delta(p_1-p_2).
$$
These vectors are defined as
$$
| p \rangle\rangle \equiv \frac{1}{\sqrt{(2\pi)^D}}\sum_{q\in {\cal O}} | q \rangle e^{i p q}.
$$
	
\end{enumerate}

\subsection{Quantum Hall effect in  condensed matter system defined on finite rectangular lattice}

We consider inhomogeneous system defined on the $D$ - dimensional lattice $\cal O$. Time remains continuous. The fermionic field $\Phi$ is defined on lattice sites. We assume that the system remains non - interacting. Keldysh Green function is defined as the following matrix
\begin{eqnarray}
	\hat{ G}(t,x|t^\prime,x^\prime)
	= -i \left(\begin{array}{cc}\langle T \Phi(t,x) \Phi^+(t^\prime,x^\prime)\rangle & -\langle  \Phi^+(t^\prime,x^\prime) \Phi(t,x)\rangle\\ \langle  \Phi(t,x) \Phi^+(t^\prime,x^\prime)\rangle & \langle \tilde{T} \Phi(t,x) \Phi^+(t^\prime,x^\prime)\rangle \end{array} \right).
	\label{KelG_S}
\end{eqnarray}
Here Heisenberg fermionic field operator $\Phi$ depends on time $t$ and spatial coordinates $x$. By $T$ we denote the time ordering while $\tilde{T}$ is anti - time ordering. $\langle ... \rangle$ means average with respect to the initial (in general, non - equilibrium) statistical ensemble.

In the following the $D+1$ dimensional vectors (with space and time components) are denoted by large Latin letters. For an operator $\hat{A}$ we denote its matrix elements by  $A(X_1,X_2) = \langle X_1 | \hat{A} | X_2 \rangle $. Since we deal with the lattice models the space components of $D+1 $ - vectors are discrete while the time components are continuous. We then define Weyl symbol of an operator $\hat A$ as the mixture of lattice Weyl symbol (with respect to discrete space components) and Wigner transformation with respect to the time component:
\begin{eqnarray}
	A_W(X|P)&=&2\int d Y^0\, \sum_{n_i = 0,1;\vec{Y} \in  {\cal O}^\prime} e^{2 \ii Y^\mu P_\mu }  A(X+Y{-}n/2,X-Y{-}n/2) \Pi_{i=1...D}\frac{{1 + {e^{-2i Y_i \pi/N}} }}{2}\frac{1+e^{2\pi i (X_i+Y_i{-} n_i/2)}}{2}, \nonumber \\ && \quad \mu =0,1,...,D.\label{WignerTr}
\end{eqnarray}
$D+1$ momentum is denoted by $P^\mu=(P^0,p)$, and $P_\mu = (P^0,-p)$. Here  $p$ is spatial momentum with $D$ components.
Keldysh Green function is an operator inverse to $\hat{\bf Q}$.
Weyl symbol of Keldysh Green function $\hat{\bf G}$ is denoted by $\hat{G}_W$, while Weyl symbol of Keldysh $\hat{\bf Q}$ is $\hat{Q}_W$.

We obtain the following results for the dynamics of lattice system written in terms of Weyl symbols of operators.

\begin{enumerate}
	\item

Weyl symbols $\hat{G}_W$ and $\hat{Q}_W$ obey Groenewold equation
\begin{equation}
	\hat{Q}_W * \hat{G}_W = 1_W.
\end{equation}
Here the Moyal product $*$ is defined as
\begin{equation}
	\left(A* B\right)(X|P) = A(X|P)\,e^{\rv{-}\ii(\overleftarrow{\partial}_{X^{\mu}}\overrightarrow{\partial}_{P_{\mu}}-\overleftarrow{\partial}_{P_{\mu}}\overrightarrow{\partial}_{X^{\mu}})/2}B(X|P).
\end{equation}

It is worth mentioning that for the complete description of the system we need the values of Weyl symbols defined on discrete spatial phase space ${\cal O}^\prime \otimes {\cal M}^\prime$. For such values of spatial momenta and coordinates the Weyl symbol of unity is equal to $1$.

\item

	We express DC conductivity (averaged over the lattice area) of the two - dimensional {\it non - interacting} systems as
	
\begin{equation}
	\sigma^{ij} =  {\frac{1}{4}} \frac{1}{(2N)^{2D}} \int \frac{dP^0}{2\pi}  \sum_{p \in   {\cal M}^\prime, x\in {\cal O}^\prime} \tr\left(\partial_{\pi_{i}}\hat{Q}_W  \left[\hat{G}_W \star \partial_{\rv{\pi_{[0}}}\hat{Q}_W  \star \partial_{\rv{\pi_{j]}}}\hat{G}_W  \right]\right)^< +{\rm c.c.}\label{MAIN}
\end{equation}
through the lesser component of expression that contains Wigner transformed Keldysh Green function $G(X|\pi)$ and its inverse $Q$ that obey Groenewold equation $Q \star G = 1$. Using representation of Keldysh Green function of Eq. (\ref{KelG_S}) the lesser component in the above representation may be rewritten explicitly as
\begin{equation}
	\sigma^{ij} =  {\frac{1}{4}} \frac{1}{(2N)^{2D}} \int \frac{dP^0}{2\pi}  \sum_{p \in   {\cal M}^\prime, x\in {\cal O}^\prime} \tr\left( \gamma^<\partial_{\pi_{i}}\hat{Q}_W  \left[\hat{G}_W \star \partial_{\rv{\pi_{[0}}}\hat{Q}_W  \star \partial_{\rv{\pi_{j]}}}\hat{G}_W  \right]\right) +{\rm c.c.}\label{MAIN}
\end{equation}
where trace is taken over Keldysh components as well as over the internal indices, while
\zz{$$
	\gamma^< = \left(\begin{array}{cc} 0 & 0 \\ 1 & 0 \end{array} \right).
	$$}
Here $(...)_{[0} (...)_{ j]} =(...)_{0} (...)_{ j} -(...)_{j} (...)_{ 0}  $ means anti-symmetrization.

\item

	We show that the above expression for the conductivity (averaged over the system area) is reduced to the following expression in case of the equilibrium system at zero temperature
$$
\bar{\sigma}^{ij}  = \frac{{\cal N}}{2\pi}\epsilon^{ij},
$$
where
\begin{eqnarray}
	{\cal N} &=& \frac{1}{3!\,}\epsilon^{\mu\nu\rho} \frac{1}{(2N)^{2D}} \int {d\Pi^3}  \sum_{p \in   {\cal M}^\prime, x\in {\cal O}^\prime}
	\tr\left(\partial_{\rv{\Pi^{\mu}}}\hat{Q}_W ^\rM  \star\hat{G}_W ^\rM \star \partial_{\rv{\Pi^{\nu}}}\hat{Q}_W ^\rM  \star\hat{G}_W ^\rM \star \partial_{\rv{\Pi^{\rho}}}\hat{Q}_W ^\rM \star \hat{G}_W ^\rM \right).
	\label{NEQ0}
\end{eqnarray}
Here the $\hat{G}_W^M$ is the  Weyl symbol of  the Matsubara Green function while $\hat{Q}_W^M$ is Weyl symbol of its inverse.
 Momentum space is Euclidean one, its points are denoted by $\Pi_i$. $\Pi^3$ is Matsubara frequency. This expression represents the infrared regularization of the one obtained in \cite{FZ2020}. {Notice that inside this expression (as well as inside Eq. (\ref{NEQ00})) the limit of infinite $N$ is to be taken in the Weyl symbols $\hat{Q}_W^{\rm M}$, $\hat{G}_W^{\rm M}$ before the differentiation with respect to momenta, while the limit of infinite lattice size in the sum over ${\cal M}^{\prime}$ and ${\cal O}^{\prime}$ is taken at the end of the calculation. Finite lattice size enters expression for $Q^M_W$ in two ways: through the coordinates that belong to the finite lattice, and directly via dependence on $N$ as on parameter. By the limit $N \to \infty$ we mean that \zg{the explicit parameter $N$ tends to infinity, while the coordinate arguments are held fixed.} Correspondingly, we should calculate $G_W^M$ as inverse with respect to Moyal product of this truncated $Q^M_W$.}

\item

One can check that Eq. (\ref{NEQ0}) is a topological invariant in the limit of infinitely large lattice when the sum over momenta is replaced by an integral. For that we need that the system was in thermal equilibrium originally, and that the thermal equilibrium corresponds to zero temperature. Moreover, we need that the Hamiltonian does not depend on time, which means that the system remains in thermal equilibrium during the whole process. The value of the average conductivity $\bar{\sigma}^{ij}$ is then robust to smooth variations of the system. (This does not refer, however, to local values of conductivity that may depend on space coordinates.) Notice, that the sum over $x$ is important for the topological invariance of this quantity.

\end{enumerate}

\section{Buot symbol and its properties }
\label{BuotSymbol}

\subsection{Definitions}

Here we consider the D - dimensional rectangular lattice with $N^D$ lattice sites. We enumerate lattice sites by variable $q = (q_1,..., q_D)$ with $q_i = 0, ..., N-1$, and denote the lattice by $\cal O$:
$$
{\cal O} = \{(m_1,...,m_D) | m_i \in \{0,1,...,N-1\}\}.
$$
 Periodic boundary conditions are chosen. Therefore, the set of lattice sites may be extended to the set ${\cal O} = Z^D/{\cal E}$ of integer numbers with equivalence ${\cal E}: q_i \equiv q_i+N$.   In turn, conjugate momenta are $p = (p_1,..., p_D)$ with $p_i = 2 \frac{\pi}{N} m_i$ with $m_i = 0, ..., N-1$.  The set of momenta is denoted by ${\cal M}$:
 $$
 {\cal M} = \{(m_1 \frac{2\pi}{N},...,m_D \frac{2\pi}{N}) | m_i \in \{0,1,...,N-1\}\}.
 $$
  The values of $m_i$ may be taken from the set of all integer numbers if equivalence relation ${\cal T}: p_i \equiv p_i + 2\pi $ is imposed. In addition, we introduce the extended lattice
$$
{\cal O}^\prime = \{(m_1,...,m_D) | m_i \in \{0,1/2,1,...,N-1/2\}\}.
$$
 And we represent it as
$$
{\cal O}^\prime =  {\cal O}\cup \tilde{\cal O}.
$$
We also define the refined momentum space
$$
{\cal M}^\prime = \{(m_1 \frac{2\pi}{N},...,m_D \frac{2\pi}{N}) | m_i \in \{0,1/2,1,...,N-1/2\}\}.
$$
 It may be represented as
$$
{\cal M}^\prime = {\cal M} \cup \tilde{\cal M}.
$$
Thus we define two operations on the lattices: ${\cal O} \to {\cal O}^\prime$ enlarges the lattice by adding points between any pair of the adjacent lattice sites. Operation ${\cal O} \to \tilde{\cal O}$ defines the lattice with the lattice sites of ${\cal O}^\prime$ that complement the points of $\cal O$. For the further convenience we also define here the  operation ${\cal O} \to {\cal O}^2$. ${\cal M} \to {\cal M}^2$:
$$
{\cal O}^2 = \{(m_1,...,m_D) | m_i \in \{0,1,...,2N-1\}\}
$$
$$
{\cal M}^2 = \{(m_1 \frac{2\pi}{N},...,m_D \frac{2\pi}{N}) | m_i \in \{0,1,...,2N-1\}\}.
$$
It doubles the lattice in all $D$ dimensions.

Hilbert space $\cal H$ of one - particle states is spanned on ket vectors
$$
| q \rangle, q \in {\cal O}.
$$
Another set of basis vectors is
$$
| p \rangle, p \in {\cal M}.
$$
We assume normalization of states
$$
\langle q | p\rangle = \frac{1}{\sqrt{N^D}} e^{i p q}.
$$
which implies completeness relations
$$
\sum_{p \in {\cal M}} |p \rangle \langle p | = \sum_{q \in {\cal O}} |q \rangle \langle q |=1.
$$

Any operator $\hat{A}$ acting in $\cal H$ may be represented as follows
\begin{equation}
	\hat{A} = \sum_{p^{\prime\prime}, p^\prime \in {\cal M}; q^{\prime\prime}, q^\prime \in {\cal O}}|p^{\prime\prime} \rangle \langle p^{\prime\prime} |q^{\prime\prime} \rangle \langle q^{\prime\prime} |\hat{A} |q^{\prime} \rangle \langle q^{\prime} |p^{\prime} \rangle \langle p^{\prime} |.
\end{equation}
The next step is an introduction of new variables $p,q,u,$ and $v$:
$$
p^{\prime\prime}=p - u, \quad p^{\prime}=p + u,
$$
$$
q^{\prime\prime}=q - v, \quad q^{\prime}=q + v.
$$
Here $p\pm u$ is assumed to be taken modulo $2\pi$ while $q\pm v$ is modulo $N$. Since
$p^{\prime}, p^{\prime\prime} \in {\cal M}$, and $q^{\prime}, q^{\prime\prime} \in {\cal O}$, in order to reproduce all possible values of these variables, we have to require that
$$
p_i+u_i \in \{0, \frac{2\pi}{N}, \frac{2\pi}{N} 2, ... , \frac{2\pi}{N}(N-1)\}
$$
while
$$q_i+v_i \in Z.$$
At the same time extra factor $1/2^2$ is to be added in order to account for the overcounting of degrees of freedom. As a result we rewrite the above expression for $\hat{A}$ as follows:
\begin{eqnarray}
	\hat{A} &=& \frac{1}{4^D}\sum_{p, u \in {\cal M}^\prime; q, v \in  {\cal O}^\prime}|p-u \rangle \langle p-u |q-v \rangle \langle q-v |\hat{A} |q+v \rangle \langle q+v |p+u \rangle \langle p+u | \nonumber\\ && \Pi_i \frac{1+e^{i 2\pi (q_i+v_i)}}{2}\frac{1+e^{i N (p_i+u_i)}}{2}.
\end{eqnarray}
Factor $\frac{1+e^{i 2\pi (q_i+v_i)}}{2}$ is introduced in order to provide that both $q_i$ and $v_i$ are either integer of half - integer. Similar, factor $\frac{1+e^{i N (p_i+u_i)}}{2}$ provides that  $
p + u \in {\cal M}
$.

\subsection{Buot symbol of operator}

We obtain the following representation for $\hat{A}$:
\begin{equation}
	\hat{A} = \frac{1}{(2N)^D}\sum_{p\in {\cal M}^\prime; q \in {\cal O}^\prime}\hat{\Delta}(p,q) A_B(p,q),
\end{equation}
where
\begin{equation}
	A_B(p,q)  = \frac{1}{2^D}\sum_{v \in {\cal O}^\prime} e^{2 i p v} \langle q-v |\hat{A} |q+v \rangle  \Pi_i\frac{1+e^{i 2\pi (q_i+v_i)}}{2},\label{Buot0}
\end{equation}
while
\begin{equation}
	\hat{\Delta}(p,q) =  \sum_{u \in {\cal M}^\prime}e^{2 i q u}|p-u \rangle  \langle p+u | \Pi_i \frac{1+e^{i N (p_i+u_i)}}{2 }.
\end{equation}
The above defined function $A_B(p,q)$ represents the lattice version of Weyl symbol of operator. For the case of the finite lattice it will be called below {\bf the Buot symbol of operator $\hat{A}$}. Notice, that this definition differs somehow from the original definition by F.Buot. However, we feel this appropriate to call this symbol in his name.

In order to understand better the properties of this symbol let us calculate the Buot symbol of unity operator:
\begin{eqnarray}
	1_B(p,q) & = & \frac{1}{2^D}\sum_{v \in {\cal O}^\prime} e^{2 i p v} \langle q-v |q+v \rangle  \Pi_i\frac{1+e^{i 2\pi (q_i+v_i)}}{2}\nonumber\\
	& = & \frac{1}{2^D}\sum_{v \in {\cal O}^\prime} e^{2 i p v} \Pi_i(\delta_{v_i,0}+\delta_{v_i,N/2}) \frac{1+e^{i 2\pi (q_i+v_i)}}{2}\nonumber\\
	& = &\Pi_i \Big(\frac{1+e^{i 2\pi q_i}}{4} +  e^{ i p_i N} \frac{1+e^{i (2\pi q_i+ \pi N)}}{4}\Big).
\end{eqnarray}
 For even $N$ we have
\begin{equation}
	1_B(p,q)= \left(\begin{array}{cc}1, & p \in {\cal M}, q \in {\cal O}\\
		0, & p \in \tilde{\cal M}, q \in \tilde{\cal O}\\0, & p \in {\cal M}, q \in \tilde{\cal O}\\0, & p \in \tilde{\cal M}, q \in {\cal O}       \end{array}\right).
\end{equation}
For odd $N$ we have
\begin{equation}
	1_B(p,q)= \left(\begin{array}{cc}1/2, & p \in {\cal M}, q \in {\cal O}\\
		-1/2, & p \in \tilde{\cal M}, q \in \tilde{\cal O}\\1/2, & p \in {\cal M}, q \in \tilde{\cal O}\\1/2, & p \in \tilde{\cal M}, q \in {\cal O}       \end{array}\right).
\end{equation}

For the operator of translation to the lattice spacing the Buot symbol is
\begin{eqnarray}
	T_j(p,q) & = & \frac{1}{2^D}\sum_{v \in {\cal O}^\prime} e^{2 i p v} \langle q-v |e^{i\hat{p}_j}|q+v \rangle  \Pi_i\frac{1+e^{i 2\pi (q_i+v_i)}}{2}\nonumber\\
	& = & \frac{1}{2^D}\sum_{v \in {\cal O}^\prime} e^{2 i p v}
	(\delta_{v_j,1/2}+\delta_{v_i,1/2+N/2}) \frac{1+e^{i 2\pi (q_j+v_j)}}{2}\nonumber\\&& \Pi_{i\ne j}(\delta_{v_i,0}+\delta_{v_i,N/2}) \frac{1+e^{i 2\pi (q_i+v_i)}}{2}\nonumber\\
	& = &\Pi_{i\ne j} \Big(\frac{1+e^{i 2\pi q_i}}{4} +  e^{ i p_i N} \frac{1+e^{i (2\pi q_i+ \pi N)}}{4}\Big)\nonumber\\&&
	\Big(e^{i p_j}\frac{1-e^{i 2\pi q_j}}{4} +  e^{ i p_j (N+1)} \frac{1-e^{i (2\pi q_j+ \pi N)}}{4}\Big).
\end{eqnarray}

Let us express the Buot symbol of operator through the matrix elements in momentum space
\begin{eqnarray}
	&&A_B(p,q)  =  \frac{1}{2^D}\sum_{p_1,p_2\in {\cal M}}\sum_{v \in {\cal O}^\prime} e^{2 i p v} \langle q-v |p_1\rangle \langle p_1 |\hat{A} |P+u\rangle \langle p_2 |p_2 \rangle  \Pi_i\frac{1+e^{i 2\pi (q_i+v_i)}}{2}\frac{1+e^{i N (P_i+u_i)}}{2}\nonumber\\ &&=
	\frac{1}{4^D}\sum_{P,u\in {\cal M}^\prime}\sum_{v \in {\cal O}^\prime} e^{2 i p v} \langle q-v |P-u\rangle \langle P-u |\hat{A} |P+u\rangle \langle P+u |q+v \rangle \Pi_i \frac{1+e^{i 2\pi (q_i+v_i)}}{2}\frac{1+e^{i N (P_i+u_i)}}{2}\nonumber\\ &&=
	\frac{1}{(4N)^D}\sum_{P,u\in {\cal M}^\prime}\sum_{v \in {\cal O}^\prime} e^{2 i p v + i (P-u)(q-v) - i (P+u)(q+v)}  \langle P-u |\hat{A} |P+u\rangle \Pi_i \frac{1+e^{i 2\pi (q_i+v_i)}}{2}\frac{1+e^{i N (P_i+u_i)}}{2}\nonumber\\ &&=
	 \frac{1}{(8N)^D}\sum_{P,u\in {\cal M}^\prime}\sum_{m_i = 0,1}\sum_{v \in {\cal O}^\prime} e^{2 i p v + i (P-u)(q-v) - i (P+u)(q+v)+i 2\pi m(q+v)}  \langle P-u |\hat{A} |P+u\rangle \Pi_i\frac{1+e^{i N (P_i+u_i)}}{2}\nonumber\\ &&=  \frac{1}{4^D}\sum_{u\in {\cal M}^\prime}\sum_{m_i = 0,1} e^{-2 i (u-\pi m) q}  \langle p+\pi m-u |\hat{A} |p+\pi m +u\rangle \Pi_i\frac{1+e^{i N (p_i+\pi m_i+u_i)}}{2}\nonumber\\ &&=
	\frac{1}{2^D}\sum_{u\in {\cal M}^\prime} e^{-2iuq}  \langle p-u |\hat{A} |p+u\rangle \Pi_i\frac{1+e^{i N (p_i+u_i)}}{2}\label{Buot}.
\end{eqnarray}
Here in the transition to the second line we change variables:
$$
p_1 = P-u, \quad p_2 = P + u
$$
$$
P = (p_1+p_2)/2, \quad u = (p_1-p_2)/2.
$$
Factor $1/2$ accounts for the overcounting of degrees of freedom during this transformation.
In the similar way we obtain
\begin{equation}
	\hat{\Delta}(p,q) = \sum_{v \in {\cal O}^\prime}e^{-2 i p v}|q-v \rangle  \langle q+v | \Pi_i \frac{1+e^{i 2\pi (q_i+v_i)}}{2 }.
\end{equation}
Notice, that in the original paper by Buot it was not specified that the sum in Eq. (\ref{Buot}) is over ${\cal M}^\prime$. Correspondingly, constraint $p+u \in {\cal M}$ resulted in function $(1+e^{i N (p_i+u_i)})/2$ was not imposed. This resulted in mistakes in the further expressions. By the present paper we correct these mistakes and introduce the proper definition of the lattice Weyl symbol. Strictly speaking our definition of function $A_B(p,q)$ differs from the original definition of Buot. We feel, however, appropriate to use this notation because we evaluate the very idea of such a formalism proposed for the first time by F.Buot.

We can express the matrix elements of an operator through the Buot symbol as follows:
\begin{eqnarray}
	\langle q_1 |\hat{A}|q_2\rangle &=& \frac{1}{(2N)^D}\sum_{p \in  {\cal M}^\prime; q \in {\cal O}^\prime}	\langle q_1 |\hat{\Delta}(p,q)|q_2\rangle  A_B(p,q)\\ &=& \sum_{p \in  {\cal M}^\prime; q \in {\cal O}^\prime}\sum_{u \in  {\cal M}^\prime}e^{2 i q u} \langle q_1 |p-u \rangle  \langle p+u | q_2\rangle  A_B(p,q)\Pi_i\frac{1+e^{i N (p_i+u_i)}}{(4 N)}	\nonumber\\ &=& \sum_{p \in {\cal M}^\prime; q \in {\cal O}^\prime}\sum_{u \in  {\cal M}^\prime}e^{ i u(2q - q_1 - q_2) + i p (q_1 - q_2) } 	 A_B(p,q)\Pi_i\frac{1+e^{i N (p_i+u_i)}}{(4 N^2)}
	\nonumber\\ &=&\frac{1}{(4N^2)^D} \sum_{p \in {\cal M}^\prime; q \in {\cal O}^\prime}\sum_{u \in  {\cal M}^\prime; n_i = 0,1}e^{ i u(2q - q_1 - q_2+N n) + i p (q_1 - q_2+N n)}	 A_B(p,q)\nonumber\\ &=&\frac{1}{(2N)^D} \sum_{p \in {\cal M}^\prime;n_i=0,1}A_B(p, (q_1 + q_2-N n)/2)e^{i p (q_1 - q_2+N n)}.	
\end{eqnarray}
Using property of the Buot symbol $A_B(p,q-N m /2)=e^{ip mN}A_B(p,q)$ (for integer - valued vector $m$) we obtain:
\begin{eqnarray}
	\langle q_1 |\hat{A}|q_2\rangle  &=&\frac{1}{N^D} \sum_{p \in {\cal M}^\prime}A_B(p,(q_1+q_2)/2)e^{ i p (q_1 - q_2) }.
\end{eqnarray}

As an example let us calculate the last row of this expression for $\hat{A}=1$ and even $N$:
\begin{eqnarray}
	&&\frac{1}{N^D} \sum_{p \in {\cal M}^\prime}1_B(p,(q_1+q_2)/2)e^{ i p (q_1 - q_2) }	\nonumber\\ & = &   \delta_{q_1,q_2}.
\end{eqnarray}
In the similar way we can derive
\begin{eqnarray}
	\langle p_1 |\hat{A}|p_2\rangle &=& \frac{1}{(2N)^D}\sum_{p \in {\cal M}^\prime; q \in {\cal O}^\prime}	\langle p_1 |\hat{\Delta}(p,q)|p_2\rangle  A_B(p,q)\\ &=& \frac{1}{(2N)^D}\sum_{p \in {\cal M}^\prime; q \in {\cal O}^\prime}\sum_{u \in  {\cal M}^\prime}e^{2 i q u} 	\langle p_1 |p-u \rangle  \langle p+u | p_2\rangle  A_B(p,q)\Pi_i \frac{1+e^{i N (p_i+u_i)}}{2}\nonumber\\ &=&
	\frac{1}{(2N)^D}\sum_{ q \in {\cal O}^\prime} \sum_{n_i=0,1} e^{2 i q ((p_2-p_1)/2+\pi n)}  A_B((p_2+p_1)/2+\pi n,q)\Pi_i \frac{1+e^{i N ((p^i_2+p^i_1)/2+\pi n_i+(p^i_2-p^i_1)/2+\pi n_i)}}{2}	\nonumber\\ &=&
	\frac{1}{(2N)^D}\sum_{ q \in {\cal O}^\prime} \sum_{n_i=0,1} e^{2 i q ((p_2-p_1)/2+\pi n )} 	 A_B((p_2+p_1)/2+\pi n ,q)\nonumber\\ &=&
	\frac{1}{N^D}\sum_{ q \in {\cal O}^\prime} e^{2 i q (p_2-p_1)/2} 	 A_B((p_2+p_1)/2,q)	.
\end{eqnarray}
The above expressions give the following two representations of the trace of an operator:
\begin{eqnarray}
	{\rm Tr}\, \hat{A} &=&\frac{1}{N^D} \sum_{p \in {\cal M}^\prime, q\in {\cal O}} A_B(p,q) = \frac{1}{N^D} \sum_{p \in {\cal M}, q\in  {\cal O}^\prime} A_B(p,q).
\end{eqnarray}
Notice that
\begin{eqnarray}
\frac{1}{N^D}\sum_{q \in   {\cal O}^\prime} A_B(p,q) &	= &
	\frac{1}{(2N)^D}\sum_{u\in {\cal M}^\prime,q \in   {\cal O}^\prime} e^{-2iuq}  \langle p-u |\hat{A} |p+u\rangle \Pi_i\frac{1+e^{i N (p_i+u_i)}}{2}\nonumber\\ & = &
	   \langle p |\hat{A} |p\rangle \Pi_i\frac{1+e^{i N p_i}}{2}
\end{eqnarray}
and
\begin{eqnarray}
	\frac{1}{N^D}\sum_{q \in   {\cal O}^\prime, p\in \tilde{\cal M}} A_B(p,q) &	= & \sum_{ p\in \tilde{\cal M}}
	\langle p |\hat{A} |p\rangle \Pi_i\frac{1+e^{i N p_i}}{2} = 0.
\end{eqnarray}
As a result we can write
\begin{eqnarray}
	{\rm Tr}\, \hat{A} &=&\frac{1}{N^D} \sum_{p \in {\cal M}^\prime, q\in {\cal O}^\prime} A_B(p,q).
\end{eqnarray}
Besides, we have
\begin{eqnarray}
	\frac{1}{N^D}\sum_{q \in  {\cal O}} A_B(p,q) &	= &
	\frac{1}{(2N)^D}\sum_{u\in {\cal M}^\prime,q \in  {\cal O}} e^{-2iuq}  \langle p-u |\hat{A} |p+u\rangle \Pi_i\frac{1+e^{i N (p_i+u_i)}}{2}\nonumber\\ & = & \frac{1}{2^D}\sum_{n_i = 0,1}
	\langle p + \pi n |\hat{A} |p + \pi n \rangle \Pi_i\frac{1+e^{i N (p_i+\pi n_i)}}{2}
\end{eqnarray}
and
\begin{eqnarray}
	\frac{1}{N^D}\sum_{q \in  {\cal O}, p \in {\cal M}} A_B(p,q)  & = & \frac{1}{2^D}\sum_{ p \in {\cal M}, n_i = 0,1}
	\langle p + \pi n |\hat{A} |p + \pi n \rangle \Pi_i\frac{1+e^{i N (p_i+\pi n_i)}}{2}.
\end{eqnarray}
Therefore, for even $N$ we have also
\begin{eqnarray}
	{\rm Tr}\, \hat{A} &=&\frac{1}{N^D} \sum_{p \in {\cal M} , q\in {\cal O}} A_B(p,q),
\end{eqnarray}
while for odd $N$:
\begin{eqnarray}
	{\rm Tr}\, \hat{A} &=&\frac{2^D}{N^D} \sum_{p \in {\cal M} , q\in {\cal O}} A_B(p,q).
\end{eqnarray}

\subsection{Star product}

Now let us consider the product of two operators $\hat{A}$ and $\hat{B}$. Our aim is to express the Buot symbol of this product through Buot symbols $A_B(p,q)$ and $B_B(p,q)$.
We have
\begin{eqnarray}
	\hat{A}\hat{B}& =&\frac{1}{(4N^2)^D} \sum_{p_1 \in  {\cal M}^\prime; q_1 \in {\cal O}^\prime}\hat{\Delta}(p_1,q_1) A_B(p_1,q_1) \sum_{p_2\in  {\cal M}^\prime; q_2 \in {\cal O}^\prime}\hat{\Delta}(p_2,q_2) A_B(p_2,q_2) \nonumber\\
	& =& \frac{1}{(4N^2)^D}\sum_{p_1,p_2 \in {\cal M}^\prime; q_1,q_2 \in {\cal O}^\prime}\hat{\Delta}(p_1,q_1)\hat{\Delta}(p_2,q_2) A_B(p_1,q_1)  A_B(p_2,q_2)
\end{eqnarray}
and
\begin{eqnarray}
	(\hat{A}\hat{B})_B(p,q)& =&\frac{1}{(8N^2)^D} \sum_{p_1,p_2,u \in  {\cal M}^\prime; q_1,q_2 \in {\cal O}^\prime}\langle p-u|\hat{\Delta}(p_1,q_1)\hat{\Delta}(p_2,q_2)|p+u\rangle \nonumber\\&& e^{-2i u q} A_B(p_1,q_1)  B_B(p_2,q_2)\Pi_i\frac{1+e^{i N (p_i+u_i)}}{2}.
\end{eqnarray}
For the matrix elements of the product of two $\Delta$ - operators we obtain:
\begin{eqnarray}
	&&	\sum_{u \in  {\cal M}^\prime}\langle p-u|\hat{\Delta}(p_1,q_1)\hat{\Delta}(p_2,q_2)|p+u\rangle e^{-2 i q u } \Pi_i\frac{1+e^{i N (p^i+u^i)}}{2} = \sum_{u,u_1,u_2 \in {\cal M}^\prime}e^{-2 i q u + 2 i q_1 u_1 + 2 i q_2 u_2}\nonumber\\&&\langle p-u|   p_1-u_1 \rangle  \langle p_1+u_1 |p_2-u_2 \rangle  \langle p_2+u_2|p+u\rangle  \Pi_i \frac{1+e^{i N (p^i_1+u^i_1)}}{2} \frac{1+e^{i N (p^i_2+u^i_2)}}{2}\frac{1+e^{i N (p^i+u^i)}}{2}	\nonumber\\&&= \sum_{n_i = 0,1}  e^{2 i( (p_2-p)(q_1-q) + (q_2-q)(p-p_1))}\Big|_{p-p_1-p_2+\pi n \in {\cal M}} e^{2\pi i n (-q + q_1+q_2)}.\nonumber
\end{eqnarray}
Here several combinations of momenta lead to the non - vanishing matrix elements like $\langle p-u|   p_1-u_1 \rangle  $:
$$
\langle p-u|   p_1-u_1 \rangle  = \sum_{m_i\in Z} \delta_{p-u,p_1 - u_1 + 2\pi m}.
$$
The constraints on the values of $u,u_1,u_2$ appear:
\begin{equation}
	\Bigl\{ \begin{array} {cc} 2\pi m & = p-u-p_1+u_1\\
		2\pi m^\prime & = p_1+u_1-p_2+u_2\\
		2\pi m^{\prime\prime} & = p_2+u_2-p-u\end{array}
\end{equation}
with integer $m_i,m_i^\prime,m_i^{\prime\prime}$.
We have solutions of these equations:
\begin{equation}
	\Bigl\{ \begin{array} {cc} u & = p_2-p_1 + \pi l \\
		u_1 & = p_2-p+\pi l\\
		u_2 & = p - p_1+\pi l \end{array}
\end{equation}
with $l_i = 0,1$.
Here the values of $u^i_1,u^i_2,u^i_3$ are taken modulo $2\pi$.

Considering analytical continuation of functions $A_B(p_1,q_1)$ and $A_B(p_2,q_2)$ to real values of arguments, we come to
\begin{eqnarray}
	8^D	(\hat{A}\hat{B})_B(p,q)& =& \sum_{n_i=0,1;p_1,p_2 \in {\cal M}^\prime; q_1,q_2 \in {\cal O}^\prime}\frac{1}{ N^{2D}} e^{2 i( (p_2-p)(q_1-q) + (q_2-q)(p-p_1))} \nonumber\\&&  A_B(p_1,q_1)  B_B(p_2,q_2)\Big|_{(p-p_1-p_2+\pi n) \in {\cal M}}  e^{2\pi i n(-q + q_1+q_2)}\nonumber\\
	& =& \sum_{p_1,\delta p_2 \in {\cal M}^\prime; q_1,\delta q_2 \in {\cal O}^\prime}\frac{1}{ N^{2D}} e^{2 i (\delta p_2 (q_1-q) + \delta q_2 (p-p_1))} \nonumber\\&&  A_B(p_1,q_1)  e^{\delta p_2\partial_p+\delta q_2 \partial_q}  B_B(p,q)\Pi_i\Big(\frac{1 + e^{i N (\delta p^i_2 - p^i_1) }}{2} + \frac{1 + e^{i N (\delta p^i_2 - p^i_1+\pi) }}{2} e^{i 2\pi (\delta q^i_2 - q^i_1) }\Big) \nonumber\\
	& =&  \sum_{p_1,\delta p_2 \in  {\cal M}^\prime; q_1,\delta q_2 \in {\cal O}^\prime}\frac{1}{ N^{2D}} e^{2 i (\delta p_2 (q_1-q) + \delta q_2 (p-p_1))}\nonumber\\&&\Pi_i\Big(\frac{1 + e^{i N (\delta p^i_2 - p^i_1) }}{2} + \frac{1 + e^{i N (\delta p^i_2 - p^i_1+\pi) }}{2} e^{i 2\pi (\delta q^i_2 - q^i_1)}\Big) \nonumber\\&&  A_B(p_1,q_1)  e^{\frac{i}{2}(\overleftarrow{\partial_q}\overrightarrow{\partial_p}-\overleftarrow{\partial_p}\overrightarrow{\partial_q})}  B_B(p,q)\label{D1__}.
\end{eqnarray}
We will consider the above expression for $p \in {\cal M}\cup {\cal M}^\prime$ and $q \in {\cal O}\cup {\cal O}^\prime$.

However, {\bf we define function $A_B(p,q)$ for any intermediate values of $p$ and $q$}:
\begin{eqnarray}
	A_B(p,q)  & =& \sum_{p_1,p_2 \in  {\cal M}^\prime; q_1,q_2 \in {\cal O}^\prime}\frac{1}{(4 N^2)^D} {e^{2 i  p_2(q_1-q) + 2 i q_2(p-p_1)}}   A_B(p_1,q_1) . \label{Bcont}
\end{eqnarray}
One can easily check the following properties of the Buot symbol {taken at the discrete values of arguments $p \in {\cal M}^{\prime}$, $q \in {\cal O}^{\prime}$ (for the analytical continuation of Eq. (\ref{Bcont}) to continuous values of arguments these properties do not hold)} ($n_i = 0,1$):
\begin{equation}
	A_B(p,q+Nn/2) = e^{- i p n N }A_B(p,q), \quad A_B(p+\pi n,q) = e^{ 2 \pi i q n}A_B(p,q)\label{Bprop}.
\end{equation}
Namely, we have for $p \in {\cal M}\cup {\cal M}^\prime; q \in {\cal O}^\prime$:
\begin{eqnarray}
	A_B(p,q+N n/2) & = & \frac{1}{2^D}\sum_{v \in {\cal O}^\prime} e^{2 i p v} \langle q+Nn/2-v |\hat{A} |q+Nn/2+v \rangle  \Pi_i\frac{1+e^{i 2\pi (q^i+Nn^i/2+v^i)}}{2}\nonumber\\& = & \frac{1}{2^D}\sum_{v \in {\cal O}^\prime} e^{2 i p (v-N n/2)} \langle q-v |\hat{A} |q+v \rangle  \Pi_i\frac{1+e^{i 2\pi (q^i+v^i)}}{2}\nonumber\\& = & \frac{1}{2^D} e^{- i p n N} \sum_{v \in {\cal O}^\prime} e^{2 i p v} \langle q-v |\hat{A} |q+v \rangle \Pi_i \frac{1+e^{i 2\pi (q^i+v^i)}}{2}\nonumber\\& = & e^{- i p n N} A_B(p,q).
\end{eqnarray}
Therefore, for continuous values of $p$ and $q$ { we obtain}:
\begin{eqnarray}
	A_B(p,q+Nn/2)  & =& \sum_{p_1,p_2 \in  {\cal M}^\prime; q_1,q_2 \in {\cal O}^\prime}\frac{1}{(4 N^2)^D} {e^{2 i  p_2(q_1-q-Nn/2) + 2 i q_2(p-p_1)}}   A_B(p_1,q_1)\nonumber\\
	& =& \sum_{p_1,p_2 \in  {\cal M}^\prime; q_1,q_2 \in {\cal O}^\prime}\frac{1}{(4 N^2)^D} {e^{2 i  p_2(q_1-q) + 2 i q_2(p-p_1)}}   A_B(p_1,q_1+Nn/2)
	\nonumber\\
	& =& \sum_{p_1,p_2 \in  {\cal M}^\prime; q_1,q_2 \in {\cal O}^\prime}\frac{1}{(4 N^2)^D} {e^{2 i  p_2(q_1-q) + 2 i q_2(p-p_1)}}  e^{- i p_1 n N} A_B(p_1,q_1) .  \label{Bcont2}
\end{eqnarray}
{For $p \in {\cal M}^\prime$, $q \in {\cal O}^\prime$ the sum over $q_2$ gives $(2N)^D \delta_{p_1 p}$, the factor $e^{- i p_1 n N}$ turns into $e^{- i p n N}$, and the periodicity property of Eq. (\ref{Bprop}) is reproduced. For continuous values of $p$ this factor remains under the sum, and the periodicity property does not hold.}

In the similar way, for discrete $p$ and $q$ we have
\begin{eqnarray}
	A_B(p+\pi n,q) & = & \frac{1}{2^D}\sum_{u \in {\cal M}^\prime} e^{-2 i q u} \langle p+\pi n-u |\hat{A} |p+\pi n+u \rangle \Pi_i \frac{1+e^{i N (p+\pi n+u)}}{2}\nonumber\\& = &\frac{1}{2^D}
	\sum_{u \in {\cal M}^\prime} e^{-2 i q (u-\pi)} \langle p-u |\hat{A} |p+u \rangle   \Pi_i \frac{1+e^{i N (p^i+u^i)}}{2}\nonumber\\& = &  e^{2 i q n \pi}  A_B(p,q).
\end{eqnarray}
{This equality is valid for the discrete values of $p$ and $q$ only.}

In Eq. (\ref{D1__}) we encounter the two sums:
The sum over $p_2$ gives:
$$
\frac{1}{(2N)^D}	\sum_{p_2 \in {\cal M}^\prime } e^{-2i p_2 (q_1 - q) } = \frac{1}{(2N)^D} \sum_{m_i = 0, 1, ..., 2N-1}e^{-2\pi i m (q_1 - q) } = \Pi_i \frac{1-e^{-\frac{4 N \pi}{N} i  (q^i_1 - q^i)}}{2N(1-e^{-\frac{2\pi}{N} i  (q^i_1 - q^i)})}= \delta_{q_1 q}
$$
for $q_1, q \in {\cal O}\cup {\cal O}^\prime$.
In the similar way the sum over $q_2$ results in
$$
\frac{1}{(2N)^D}	\sum_{q_2 \in  {\cal O}^\prime} e^{2i q_2 (p - p_1) } = \delta_{p_1 p}
$$
for $p_1, p \in  {\cal M}^\prime$.
We recall that in Eq. (\ref{D1}) the arguments of $A_B, B_B$ are real - valued, and according to the definition of the Buot symbol we obtain
\begin{eqnarray}
	8^D	(\hat{A}\hat{B})_B(p,q)
	&=&  \sum_{n_i,m_i = 0,1; p_1,\delta p_2 \in  {\cal M}^\prime; q_1,\delta q_2 \in {\cal O}^\prime}\frac{1}{2^D N^{2D}} e^{2 i (\delta p_2 (q_1-q) + \delta q_2 (p-p_1))}\nonumber\\&&\Big(  e^{-i N  n (p_1-\pi m) }A_B(p_1+\pi m,q_1-N n/2))  e^{- 2\pi i q_1 } \Big) \nonumber\\&&   e^{\frac{i}{2}(\overleftarrow{\partial_q}\overrightarrow{\partial_p}-\overleftarrow{\partial_p}\overrightarrow{\partial_q})}  B_B(p,q).\label{D1}
\end{eqnarray}
Using Eq. (\ref{Bprop}) and Eq. (\ref{Bcont}) we obtain
\begin{eqnarray}
	&&	(\hat{A}\hat{B})_B(p,q)\Big|_{p\in {\cal M}^\prime,q \in {\cal O}^\prime}
	=  A_B(p,q)  e^{\frac{i}{2}(\overleftarrow{\partial_q}\overrightarrow{\partial_p}-\overleftarrow{\partial_p}\overrightarrow{\partial_q})}  B_B(p,q).
\end{eqnarray}

\subsection{Alternative derivation of star identity}
Let us give the alternative derivation of this equality. It is assumed that $p \in {\cal M}^\prime$ while $q \in {\cal O}^\prime$. We start from the star product of Buot symbols and come back to the Buot symbol of the product as follows.
\begin{eqnarray}
	&&A_B(p,q)  e^{\frac{i}{2}(\overleftarrow{\partial_q}\overrightarrow{\partial_p}-\overleftarrow{\partial_p}\overrightarrow{\partial_q})}  B_B(p,q)  = \nonumber\\
	&& =  \sum_{p_1,\delta p_2 \in  {\cal M}^\prime; q_1,\delta q_2 \in {\cal O}^\prime}\frac{1}{ (2N)^{2D}} e^{2 i (\delta p_2 (q_1-q) + \delta q_2 (p-p_1))}  A_B(p_1,q_1)  e^{\frac{i}{2}(\overleftarrow{\partial_q}\overrightarrow{\partial_p}-\overleftarrow{\partial_p}\overrightarrow{\partial_q})}  B_B(p,q)\nonumber\\&&=
	 \sum_{p_1,p_2 \in {\cal M}^\prime; q_1,q_2 \in {\cal O}^\prime}\frac{1}{ (2N)^{2D}} e^{2 i( (p_2-p)(q_1-q) + (q_2-q)(p-p_1))}   A_B(p_1,q_1)  B_B(p_2,q_2)\nonumber\\
	 &&=
	 \sum_{p_1,p_2 \in {\cal M}^\prime; v_1,v_2,q_1,q_2 \in {\cal O}^\prime}\frac{1}{ 2^{2D}(2N)^{2D}} e^{2 i( (p_2-p)(q_1-q) + (q_2-q)(p-p_1))}  \zg{e^{2 i p_1 v_1+2 i p_2v_2}} \langle q_1-v_1 |\hat{A} |q_1+v_1 \rangle\langle q_2-v_2 |\hat{B} |q_2+v_2 \rangle\Big|_{q_1+v_1,q_2+v_2\in {\cal O}}  \nonumber\\
	 &&=
	 \sum_{v_1,v_2
	 	 \in {\cal O}^\prime}\frac{1}{ 2^{2D}} e^{2 i p (v_1+v_2)}   \langle q-v_1-v_2 |\hat{A} |q+v_1-v_2 \rangle\langle q +v_1-v_2 |\hat{B} |q+v_1+v_2 \rangle\Big|_{q+v_1+v_2,q+v_1-v_2\in {\cal O}}
 	 \nonumber\\
 	 &&=
 	 \sum_{v_+,v_-
 	  \in {\cal O}^{2\prime}}\frac{1}{ 2^{3D}} e^{2 i p v_+}   \langle q-v_+ |\hat{A} |q+v_- \rangle\langle q +v_- |\hat{B} |q+v_+ \rangle\Big|_{q+v_+,q+v_-,v_++v_-\in {\cal O}}
  	 \nonumber\\
  	 &&=
  	 \sum_{v_+,v_-
  	 	 \in {\cal O}^{\prime}}\frac{1}{ 2^{D}} e^{2 i p v_+}   \langle q-v_+ |\hat{A} |q+v_- \rangle\langle q +v_- |\hat{B} |q+v_+ \rangle\Big|_{q+v_+,q+v_-,v_++v_-\in {\cal O}}
   	\nonumber\\
   	&&=
   	\sum_{n_i=0,1;v_+,v_-
   		 \in {\cal O}^{\prime}}\frac{1}{ 2^{2D}} e^{2 i p v_+ + 2\pi i n (v_++v_-)}   \langle q-v_+ |\hat{A} |q+v_- \rangle\langle q +v_- |\hat{B} |q+v_+ \rangle\Big|_{q+v_+,q+v_-\in {\cal O}} .
\label{D1__2}
\end{eqnarray}
One can see that for $q_i \in Z$ both $v_+^i$ and $v_-^i$ are integer. At the same time if $q_i$ is half integer, both $v_+^i$ and $v_-^i$ are half integer. In both cases $v_+ + v_-$ is integer. As a result
\begin{eqnarray}
	&&A_B(p,q)  e^{\frac{i}{2}(\overleftarrow{\partial_q}\overrightarrow{\partial_p}-\overleftarrow{\partial_p}\overrightarrow{\partial_q})}  B_B(p,q) = \nonumber\\
	&&=
	\sum_{n=0,1;v_+,v_-
		\in {\cal O}^{\prime}}\frac{1}{ 2^{2D}} e^{2 i p v_+ + 2\pi i n (v_++v_-)}   \langle q-v_+ |\hat{A} |q+v_- \rangle\langle q +v_- |\hat{B} |q+v_+ \rangle\Big|_{q+v_+,q+v_-\in {\cal O}}\nonumber\\
	&&=
\sum_{v_+,v_-
\in {\cal O}^\prime}\frac{1}{ 2^{D}} e^{2 i p v_+ }    \langle q-v_+ |\hat{A} |q+v_- \rangle\langle q +v_- |\hat{B} |q+v_+ \rangle\Big|_{q+v_+,q+v_-\in {\cal O}}\nonumber\\
&&=
\sum_{v_+
\in {\cal O}^\prime}\frac{1}{ 2^{D}} e^{2 i p v_+ }   \langle q-v_+ |\hat{A} \hat{B} |q+v_+ \rangle = (\hat{A} \hat{B})_B(p,q).
	\label{D1__2}
\end{eqnarray}

\subsection{Trace of the product}

Let us consider the trace of the product of two operators $\hat{A}$ and $\hat{B}$.
We have
\begin{eqnarray}
	\hat{A}\hat{B}& =&\frac{1}{(4N^2)^D} \sum_{p_1 \in  {\cal M}^\prime; q_1 \in {\cal O}^\prime}\hat{\Delta}(p_1,q_1) A_B(p_1,q_1) \sum_{p_2\in {\cal M}\cup {\cal M}^\prime; q_2 \in {\cal O}^\prime}\hat{\Delta}(p_2,q_2) A_B(p_2,q_2) \nonumber\\
	& =& \frac{1}{(4N^2)^D}\sum_{p_1,p_2 \in  {\cal M}^\prime; q_1,q_2 \in {\cal O}^\prime}\hat{\Delta}(p_1,q_1)\hat{\Delta}(p_2,q_2) A_B(p_1,q_1)  A_B(p_2,q_2)
\end{eqnarray}
and
\begin{eqnarray}
	{\rm Tr}(\hat{A}\hat{B})& =&\frac{1}{(4N^2)^D} \sum_{p\in {\cal M}}\sum_{p_1,p_2 \in  {\cal M}^\prime; q_1,q_2 \in {\cal O}^\prime}\langle p|\hat{\Delta}(p_1,q_1)\hat{\Delta}(p_2,q_2)|p\rangle  A_B(p_1,q_1)  B_B(p_2,q_2).
\end{eqnarray}
For the matrix elements of the product of two $\Delta$ - operators we obtain:
\begin{eqnarray}
	&&	\sum_{p \in {\cal M}}\langle p|\hat{\Delta}(p_1,q_1)\hat{\Delta}(p_2,q_2)|p\rangle  = \sum_{p\in{\cal M}; u_1,u_2 \in {\cal M}^\prime}e^{2 i q_1 u_1 + 2 i q_2 u_2}\nonumber\\&&\langle p|   p_1-u_1 \rangle  \langle p_1+u_1 |p_2-u_2 \rangle  \langle p_2+u_2|p\rangle  \sum_i\frac{1+e^{i N (p^i_1+u^i_1)}}{2} \frac{1+e^{i N (p^i_2+u^i_2)}}{2}	\nonumber\\&&= N^D \sum_{n_i,m_i = 0,1}  \delta_{p_1 -  p_2 + \pi n,0}\delta_{q_1 - q_2-N m/2,0} e^{2\pi i q_1 n + i N  m p_1 - i \pi nm N}.
\end{eqnarray}
Here several combinations of momenta lead to the nonvanishing matrix elements like $\langle p|   p_1-u_1 \rangle  $:
$$
\langle p|   p_1-u_1 \rangle  = \sum_{m_i\in Z} \delta_{p,p_1 - u_1 + 2\pi m}.
$$
The constraints on the values of $p,u_1,u_2$ appear:
\begin{equation}
	\Bigl\{ \begin{array} {cc} 2\pi m & = p-p_1+u_1\\
		2\pi m^\prime & = p_1+u_1-p_2+u_2\\
		2\pi m^{\prime\prime} & = p_2+u_2-p\end{array}
\end{equation}
with integer $m,m^\prime,m^{\prime\prime}$.
We have the following solutions of these equations:
\begin{equation}
	\Bigl\{ \begin{array} {cc} u_2 & = p-p_2\\
		u_1 & = p_1 - p \\
		p_1 & = p_2 + \pi n \end{array}
\end{equation}
with $n_i = 0,1$.
Here all values of variables are taken modulo $2\pi$.

We are in the position to obtain the necessary expression for the trace of product:
\begin{eqnarray}
	{\rm Tr}(\hat{A}\hat{B})&=&\frac{1}{(4N)^D} \sum_{p_1 \in{\cal M}^\prime; q_1 \in {\cal O}^\prime; n_i,m_i = 0,1} \Big(A_B(p_1,q_1)B_B(p_1+\pi n,q_1-N m /2) e^{2\pi i q n + i N p_1 m -iN nm\pi}
	\Big)\nonumber\\& = &  \frac{1}{N^D} \sum_{p_1 \in  {\cal M}^\prime; q_1 \in {\cal O}^\prime}A_B(p_1,q_1)B_B(p_1,q_1).
\end{eqnarray}

\subsection{Summary of the properties of the Buot symbol}

For $q\in {\cal O}^\prime, p \in {\cal M}^\prime  $ we define
\begin{eqnarray}
	A_B(p,q) & = & \frac{1}{2^D}\sum_{v \in {\cal O}^\prime} e^{2 i p v} \langle q-v |\hat{A} |q+v \rangle  \Pi_i\frac{1+e^{i 2\pi (q_i+v_i)}}{2}\nonumber\\ \nonumber\\ &=&
	\frac{1}{2^D}\sum_{u\in {\cal M}^\prime} e^{-2iuq}  \langle p-u |\hat{A} |p+u\rangle \Pi_i\frac{1+e^{i N (p_i+u_i)}}{2}.
\end{eqnarray}
This definition is extended further to any values of $p,q \in R^D$ as follows:
\begin{eqnarray}
	A_B(p,q)  & =& \sum_{p_1,p_2 \in  {\cal M}^\prime; q_1,q_2 \in {\cal O}^\prime}\frac{1}{(4 N^2)^D} {e^{2 i  p_2(q_1-q) + 2 i q_2(p-p_1)}}   A_B(p_1,q_1) .\label{Bcont0}
\end{eqnarray}

The matrix elements of operator $\hat A$ are expressed through the Buot symbol of this operator as follows:
\begin{eqnarray}
	\langle q_1 |\hat{A}|q_2\rangle  &=&\frac{1}{N^D} \sum_{p \in {\cal M}^\prime}A_B(p,(q_1+q_2)/2)e^{ i p (q_1 - q_2) }
\end{eqnarray}
and
\begin{eqnarray}
	\langle p_1 |\hat{A}|p_2\rangle &=&
	\frac{1}{N^D}\sum_{ q \in {\cal O}^\prime} e^{ i q (p_2-p_1)} 	 A_B((p_2+p_1)/2,q).	
\end{eqnarray}

We formulate the following properties of the Buot symbol:

\begin{enumerate}
	\item
	
	Trace property.
	$$	
	{\rm Tr}\, \hat{A} =\frac{1}{N^D} \sum_{p \in {\cal M}^\prime, q\in {\cal O}} A_B(p,q) = \frac{1}{N^D} \sum_{p \in {\cal M}, q\in {\cal O}^\prime} A_B(p,q)= \frac{1}{N^D} \sum_{p \in  {\cal M}^\prime, q\in  {\cal O}^\prime} A_B(p,q)
	$$	
	
	\item

	Second trace identity.
	
	$$	
	{\rm Tr} \hat{A}\hat{B} = \frac{1}{N^D} \sum_{p \in  {\cal M}^\prime, q\in {\cal O}^\prime} A_B(p,q) B_B(p,q)
	$$	
	
	\item
	
	Periodicity.

	$$
	A_B(p,q+N n/2) = e^{- i p N n}A_B(p,q), \quad A_B(p+\pi n,q) = e^{ 2 \pi i q n}A_B(p,q)
	$$
	with $n_i = 0,1$. {(These properties are valid for the discrete values of arguments $p \in {\cal M}^{\prime}$, $q \in {\cal O}^{\prime}$ only.)}
	
	\item Star property.
	
	\begin{eqnarray}
		&&	(\hat{A}\hat{B})_B(p,q)\Big|_{p \in  {\cal M}^\prime, q\in {\cal O}^\prime}
		=  A_B(p,q)  e^{\frac{i}{2}(\overleftarrow{\partial_q}\overrightarrow{\partial_p}-\overleftarrow{\partial_p}\overrightarrow{\partial_q})}  B_B(p,q)
	\end{eqnarray}
	
	\item Buot symbol of unity.
	
	$$
	1_B(p,q)  = \Pi_i \Big(\frac{1+e^{i 2\pi q_i}}{4} +  e^{ i p_i N} \frac{1+e^{i (2\pi q_i+ \pi N)}}{4}\Big)
	$$
	
\end{enumerate}

\section{Modifications of the Buot symbol of operator, and the definition of lattice Weyl symbol}
\label{ModifBuotSymbol}

\subsection{The $1D$ construction that uses auxiliary lattice}

First let us consider for simplicity the case of the one - dimensional system. In the following we will consider the even values of $N$, and impose the following  constraints on all operators.
$$
\langle p_1 | \hat{A} | p_2 \rangle = \langle p_1 + 2 \pi/N | \hat{A} | p_2 + 2\pi/N \rangle$$
for $p_1, p_2 \in {\cal M}$, while
$$ \langle p_1 | \hat{A} | p_2 \rangle = 0
$$
for  even $N p_1/(2\pi)$ and odd $N p_2/(2\pi)$.

Now $\cal M$ may be divided into the two pieces ${\cal M}  = {\cal M}_{1}\cup {\cal M}_{2}$. Here
$${\cal M}_1 = \{0,2 \frac{2\pi}{N},4 \frac{2\pi}{N}, ..., (N-2) \frac{2\pi}{N}\}$$
while
$${\cal M}_2 = \{1 \frac{2\pi}{N},3 \frac{2\pi}{N} ,5 \frac{2\pi}{N}, ..., (N-1) \frac{2\pi}{N}\}.$$
Notice that ${\cal M}^\prime_{1} = {\cal M}$ while ${\cal M}_{2} = \tilde{\cal M}_1$.
By $\hat{A}_1$ we denote restriction of operator $\hat{A}$ to space spanned on vectors $|p\rangle$ with $ p \in {\cal M}_1$.

Hilbert space ${\cal H}$ of one - particle states on the reduced momentum lattice ${\cal M}_1$ is spanned on ket vectors
$$
| p \rangle_1, p \in {\cal M}_1.
$$
Another set of basis vectors is
$$
| q \rangle_1, q \in {\cal O}^{1/2}
$$
with
$${\cal O}^{1/2} = \{0, 1, 2, ..., (N/2-1)\}.$$
We also use
$${\cal O}^{1/2 \prime} = \{1/2,1, 3/2 , ..., (N/2-1/2)\}.$$
Here
$$
| q \rangle_1 = \frac{1}{\sqrt{N/2}}\sum_{p\in {\cal M}_1} |p \rangle_1 e^{-i p q}.
$$
These vectors are defined also for any $q \in {\cal O}$. We assume normalization of states
$$
_1\langle q | p\rangle_1 = \frac{1}{\sqrt{N/2}} e^{i p q}
$$
which implies completeness relations within ${\cal H}_1$:
$$
\sum_{p \in {\cal M}_1} |p \rangle_1   {}_1\langle p | = \sum_{q \in {\cal O}^{1/2}} |q \rangle_1   {}_1 \langle q |=1.
$$
In the similar way the Hilbert space ${\cal H}_2$ for ${\cal M}_2 $ is defined:
$$
| q \rangle_2, q \in {\cal O}^{1/2}.
$$
Another set of basis vectors is
$$
| p \rangle_2, p \in {\cal M}_2
$$
with
$${\cal O}^{1/2} = \{0, 1, 2, ..., (N/2-1)\}.$$
One can see that $({\cal O}^{1/2 \prime})^2 = {\cal O}^\prime$ and  $({\cal O}^{1/2})^2 = {\cal O}$.
Here
$$
| q \rangle_2 = \frac{1}{\sqrt{N/2}}\sum_{p\in {\cal M}_2} | p \rangle_2 e^{-i p q}
$$
We assume normalization of states
$$
_2\langle q | p\rangle_2 = \frac{1}{\sqrt{N/2}} e^{i p q},
$$
which implies completeness relations in ${\cal H}_2$:
$$
\sum_{p \in {\cal M}_2} |p \rangle_2   {}_2\langle p | = \sum_{q \in {\cal O}^{1/2}} |q \rangle_2   {}_2 \langle q |=1.
$$

One can consider inclusion of ${\cal H}_1$ into ${\cal H}$ with the following rules:
$$
| p \rangle_1 = | p \rangle, \quad | q \rangle = \frac{| q \rangle_1 + | q \rangle_2}{\sqrt{2}}.
$$
Notice that not all vectors $|q\rangle_1 $ for $q \in {\cal O}$ are independent:
$$
| q + N/2 \rangle_1 = | q \rangle_1.
$$
Therefore, we may choose as basis vectors those with $q\in {\cal O}^{1/2}$. In the similar way
$$
| q + N/2 \rangle_2 = -| q \rangle_2,
$$
while
$$
| q + N \rangle = | q \rangle.
$$

Modified Buot symbol of operator $\hat{A}_1$ is defined as the Buot symbol of operator $\hat{A}$:

\begin{eqnarray}
	A_{1,{\cal B}}(p,q) &\equiv & A_B(p,q)  = \frac{1}{2}\sum_{v \in  {\cal O}^\prime } e^{2 i p v} \langle q-v |\hat{A} |q+v \rangle  \frac{1+e^{i 2\pi (q+v)}}{2}
	\nonumber\\ & = & \frac{1}{4} \sum_{v \in  {\cal O}^\prime } e^{2 i p v} \,
	_1\langle q-v |\hat{A}_1 |q+v \rangle_1 ({1 + e^{-4i v \pi/N }})\frac{1+e^{i 2\pi (q+v)}}{2} 	\nonumber\\ & = & \frac{1}{2} \sum_{v \in  {\cal O}_1^\prime } e^{2 i p v} \,
	_1\langle q-v |\hat{A}_1 |q+v \rangle_1 ({1 + e^{-4i v \pi/N }})\frac{1+e^{i 2\pi (q+v)}}{2}\frac{1+e^{i p N}}{2}  \label{Weyl}
\end{eqnarray}
for $q\in  {\cal O}^\prime $, $p\in   {\cal M}^\prime$. Factor $\frac{1+e^{i N p}}{2}$ entering this expression results in vanishing of the modified Buot symbol for $p \in {\cal M}^\prime \setminus {\cal M}$.

Here we express matrix elements of operator $\hat{A}$ as follows:
\begin{eqnarray}
	\langle q_1 |\hat{A} |q_2 \rangle &=& \frac{1}{N} \sum_{p_1,p_2 \in {\cal M}_1 \cup {\cal M}_2}  \langle p_1 |\hat{A} |p_2 \rangle e^{-i p_2 q_2 + i p_1 q_1}\nonumber\\
	&=& \frac{1}{N} \sum_{p_1,p_2 \in {\cal M}_1 } \Big(e^{-i p_2 q_2 + i p_1 q_1} \langle p_1 |\hat{A} |p_2 \rangle + e^{-i (p_2 + 2\pi/N) q_2 + i (p_1 + 2\pi/N) q_1} \langle p_1+2\pi/N |\hat{A} |p_2+2\pi/N \rangle\Big)
	\nonumber\\
	&=& \frac{2}{N} \sum_{p_1,p_2 \in {\cal M}_1 } e^{-i p_2 q_2 + i p_1 q_1} \langle p_1 |\hat{A} |p_2 \rangle \frac{1 + e^{-i (q_2  -  q_1) 2\pi/N }}{2}
	\nonumber\\
	&=& _1\langle q_1 |\hat{A}_1 |q_2 \rangle_1 \frac{1 + e^{-i (q_2  - q_1) 2\pi/N }}{2}.
\end{eqnarray}
This expression works both for $q_1,q_2 \in {\cal O}^{1/2}$ and for $q_1,q_2 \in {\cal O}$ assuming $|q + N/2\rangle_1 = |q \rangle_1$.

This definition may be easily extended to the whole range of real values of $p$ and $q$ as follows:
\begin{eqnarray}
	A_{\cal B}(p,q)  & = & \sum_{p_2 \in {\cal M}^\prime; q_2 \in  {\cal O}^\prime;p_1 \in {\cal M}^\prime; q_1 \in  {\cal O}^\prime}\frac{1}{4 N^2} {e^{2 i  p_2(q_1-q) + 2 i q_2(p-p_1)}}   A_{1,{\cal B}}(p_1,q_1) .
\end{eqnarray}
With this extension we obtain the basic property of the modified Buot symbol that follows directly from the corresponding property of Buot symbol:
\begin{eqnarray}
	&&	(\hat{A}\hat{B})_{\cal B}(p,q)\Big|_{p\in {\cal M}^\prime ,q \in {\cal O}^\prime}
	=  A_{\cal B}(p,q)  e^{\frac{i}{2}(\overleftarrow{\partial_q}\overrightarrow{\partial_p}-\overleftarrow{\partial_p}\overrightarrow{\partial_q})}  B_{\cal B}(p,q).
\end{eqnarray}

\subsection{The $D$ - dimensional construction with auxiliary lattice}

Now we extend the above construction to the case of the $D$ - dimensional system. We still consider the even values of $N$, and impose the following  constraints on all operators:
$$
\langle p_1 | \hat{A} | p_2 \rangle = \langle p_1 + 2 \sum_j(\pi/N) n^j e_j | \hat{A} | p_2 + 2\sum_j (\pi/N) n^j e_j \rangle,$$
where $e_j = (0,...,1,...0)$ is unity vector in the $j$ - th direction while $n_j = 0,1$.
Here $p_1, p_2 \in {\cal M}$.
We define
$$
{\cal M}_1 = \{(m_1 \frac{2\pi}{N},...,m_D \frac{2\pi}{N}) | m_i \in \{0,2,4,...,N-2\}\}.
$$
We also require
$$ \langle p_1 | \hat{A} | p_2 +  2(\pi/N) e_j\rangle = 0
$$
for  $p_1,p_2 \in {\cal M}_1$ and any $j$.

Now $\cal M$ may be divided into the two pieces ${\cal M}  = {\cal M}_1\cup {\cal M}_2$. Here
 ${\cal M}_2$ is the remaining piece of $\cal M$.

By $\hat{A}_1$ we denote restriction of operator $\hat{A}$ to space spanned on vectors $|p\rangle$ with $ p \in {\cal M}_1$.

Hilbert space ${\cal H}$ of one - particle states on the reduced momentum lattice ${\cal M}_1$ is spanned on ket vectors
$$
| p \rangle_1, p \in {\cal M}_1.
$$
Another set of basis vectors is
$$
| q \rangle_1, q \in {\cal O}^{1/2}
$$
with
$$
{\cal O}^{1/2} = \{(m_1,...,m_D) | m_i \in \{0,1,2,...,N/2-1\}\}.
$$
We also have
$$
{\cal O}^{1/2 \prime} = \{(m_1,...,m_D) | m_i \in \{0,1/2,1,...,N/2-1/2\}\}.
$$

Here
$$
| p \rangle_1 = \frac{1}{\sqrt{(N/2)^D}}\sum_{x\in {\cal O}_1} | x \rangle_1 e^{i p x}.
$$
We assume normalization of states
$$
_1\langle x | p\rangle_1 = \frac{1}{\sqrt{(N/2)^D}} e^{i p x},
$$
which implies completeness relations
$$
\sum_{p \in {\cal M}_1} |p \rangle_1   {}_1\langle p | = \sum_{q \in {\cal O}_1} |q \rangle_1   {}_1 \langle q |=1.
$$

One can consider inclusion of ${\cal H}_1$ into ${\cal H}$ with the following rules:
$$
| p \rangle_1 = | p \rangle.
$$
Notice that
$$
| q + e_j N/2 \rangle_1 = | q \rangle_1,
$$
while
$$
| q + e_j N \rangle = | q \rangle.
$$

Modified Buot symbol of operator $\hat{A}_1$ is defined as the Buot symbol of operator $\hat{A}$:

\begin{eqnarray}
	A_{1,{\cal B}}(p,q) &\equiv & A_B(p,q)  = \frac{1}{2^D}\sum_{v \in  {\cal O}^\prime } e^{2 i p v} \langle q-v |\hat{A} |q+v \rangle \Pi_i \frac{1+e^{i 2\pi (q_i+v_i)}}{2} \nonumber\\  & = & \frac{1}{2^D} \sum_{v \in  {\cal O}^{1/2 \prime} } e^{2 i p v} \,
	_1\langle q-v |\hat{A}_1 |q+v \rangle_1 \Pi_i({1 + e^{-4i v_i \pi/N }})\frac{1+e^{i 2\pi (q_i+v_i)}}{2}\frac{1+e^{i N p_i }}{2} \label{Weyl}
\end{eqnarray}
for $q\in  {\cal O}^\prime$, $p\in   {\cal M}^\prime$. Again, the modified Buot symbol is vanisning for $p \in {\cal M}^\prime \setminus {\cal M}$.

Here we express matrix elements of operator $\hat{A}$ as follows:
\begin{eqnarray}
	\langle q_1 |\hat{A} |q_2 \rangle &=& \frac{1}{N^D} \sum_{p_1,p_2 \in {\cal M}}  \langle p_1 |\hat{A} |p_2 \rangle e^{-i p_2 q_2 + i p_1 q_1}\nonumber\\
	&=& \frac{1}{N^D} \sum_{n_i = 0,1;p_1,p_2 \in {\cal M}_1 } \Big( e^{-i (p_2 + 2\pi n/N) q_2 + i (p_1 + 2\pi n/N) q_1} \langle p_1+2\pi n/N |\hat{A} |p_2+2\pi n/N \rangle\Big)
	\nonumber\\
	&=& \frac{2^D}{N^D} \sum_{p_1,p_2 \in {\cal M}_1 } e^{-i p_2 q_2 + i p_1 q_1} \langle p_1 |\hat{A} |p_2 \rangle \Pi_i\frac{1 + e^{-i (q^i_2  -  q^i_1) 2\pi/N }}{2}
	\nonumber\\
	&=& _1\langle q_1 |\hat{A}_1 |q_2 \rangle_1 \Pi_i\frac{1 + e^{-i (q^i_2  -  q^i_1) 2\pi/N }}{2}.
\end{eqnarray}


This definition may be easily extended to the whole range of real values of $p$ and $q$ as follows:
\begin{eqnarray}
	A_{\cal B}(p,q)  & = & \sum_{p_2 \in {\cal M}^\prime; q_2 \in  {\cal O}^\prime;p_1 \in {\cal M}^\prime; q_1 \in  {\cal O}^\prime}\frac{1}{(4 N^2)^D} {e^{2 i  p_2(q_1-q) + 2 i q_2(p-p_1)}}   A_{1,{\cal B}}(p_1,q_1) .
\end{eqnarray}
With this extension we obtain the basic property of the modified Buot symbol that follows directly from the corresponding property of Buot symbol:
\begin{eqnarray}
	&&	(\hat{A}\hat{B})_{\cal B}(p,q)\Big|_{p\in {\cal M}^\prime ,q \in {\cal O}^\prime}
	=  A_{\cal B}(p,q)  e^{\frac{i}{2}(\overleftarrow{\partial_q}\overrightarrow{\partial_p}-\overleftarrow{\partial_p}\overrightarrow{\partial_q})}  B_{\cal B}(p,q).
\end{eqnarray}

Let us rewrite the definition of the modified Buot symbol as follows
\begin{eqnarray}
&&	A_{1, {\cal B}}(p,q)  =
	\frac{1}{2^D}\sum_{u\in  {\cal M}^\prime} e^{-2iuq}  \langle p-u |\hat{A} |p+u\rangle \Pi_i\frac{1+e^{i N (p_i+u_i)}}{2}\nonumber\\ & = &
	\frac{1}{2^D}\sum_{n_i = 0,1;u\in  {\cal M}^\prime} e^{-2iuq}  \, _1\langle p-u -2\pi n/N  |\hat{A} |p+u -2\pi n/N\rangle_1 \Pi_i\frac{1+e^{i N (p_i+u_i)}}{2}\frac{1+e^{i N (p_i+u_i - 2\pi n/N)/2}}{2}.
\end{eqnarray}

For the operator of translation to the lattice spacing the modified Buot symbol is
\begin{eqnarray}
	T_j(p,q) & = & \frac{1}{2^D}\sum_{v \in  {\cal O}_1^\prime} e^{2 i p v}\, _1\langle q-v |e^{i\hat{p}_j}|q+v \rangle_1  \Pi_i ({1 + e^{-4i v_i \pi/N }}) \frac{1+e^{i 2\pi (q_i+v_i)}}{2}\frac{1+e^{i N p_i}}{2}\nonumber\\
	& = & \frac{1}{2^D}\sum_{v \in  {\cal O}_1^\prime} e^{2 i p v}
	(\delta_{v_j,1/2}+\delta_{v_j,1/2+N/4}) ({1 + e^{-4i v_j \pi/N }})\frac{1+e^{i 2\pi (q_j+v_j)}}{2}\frac{1+e^{i N p_j}}{2}\nonumber\\&& \Pi_{i\ne j}(\delta_{v_i,0}+\delta_{v_i,N/4})({1 + e^{-4i v_i \pi/N }}) \frac{1+e^{i 2\pi (q_i+v_i)}}{2}\frac{1+e^{i N p_i}}{2}\nonumber\\
	& = &e^{i p_j}\Bigl(\frac{1-e^{i 2\pi q_i}}{2} \frac{1+e^{-i 2\pi/N}}{2}+e^{ipN/2}\frac{1-e^{i 2\pi q_i+i\pi N/2}}{2} \frac{1-e^{-i 2\pi/N}}{2} \Bigr) \frac{1+e^{i N p_j}}{2}\nonumber\\&& \Pi_{i\ne j} \Big(\frac{1+e^{i 2\pi q_i}}{2} \Big)\frac{1+e^{i N p_i}}{2}.
\end{eqnarray}

\subsection{Construction without auxiliary lattice of modified Buot symbol. The doubly modified Buot symbol}

The above given construction gives expression for the modified Buot symbol for the operator defined on the lattice with lattice spacing equal to $1$ and the number of lattice sites $(N/2)^D$. In order to build the modified Buot symbol for the lattice with unit lattice spacing and the number of lattice sites equal to $N^D$ we need to substitute $N \to 2N$. In the definition of momentum space we omit the subscript $1$, that is we substitute ${\cal M}_1 \to {\cal M}$:
$$
{\cal M} = \{(m_1 \frac{2\pi}{N},...,m_D \frac{2\pi}{N}) | m_i \in \{0,1,2,...,N-1\}\}
$$
while coordinate space is $\cal O$ in place of ${\cal O}^{1/2}$:
$$
{\cal O} = \{(m_1 ,...,m_D) | m_i \in \{0,1,2,...,N-1\}\}.
$$
Correspondingly, the previous ${\cal O}$ becomes ${\cal O}^2$ while ${\cal M}$ becomes ${\cal M}^\prime$. Overall we have the following change of notations:
$$
N \to 2 N, \quad {\cal O}^{1/2} \to {\cal O}, \quad {\cal O} \to {\cal O}^2, \quad {\cal M}\to {\cal M}^\prime, \quad {\cal M}^\prime \to {\cal M}^{\prime \prime}.
$$
The modified Buot symbol is given by
\begin{eqnarray}
	A_{{\cal B}}(p,q) & = &  \frac{1}{2^D} \sum_{v \in  {\cal O}^\prime } e^{2 i p v} \,
	\langle q-v |\hat{A} |q+v \rangle \Pi_i({1 + {e^{2i v_i \pi/N}} })\frac{1+e^{i 2\pi (q_i+v_i)}}{2}\frac{1+e^{2i N p_i}}{2}\nonumber\\
	& = & \frac{1}{2^D} \sum_{n_i = 0,1;  p_- \in {\cal M}^{\prime \prime}   }  \,
	\langle p + \pi n/N - p_-|\hat{A}|p + \pi n/N + p_- \rangle \,  e^{-2 i p_- q}\Pi_i \frac{1+e^{i 2 N (p_i+p_-^i)}}{2}\frac{1+e^{Ni (p^i+  \pi n /N+ p_-^i)}}{2}
	\nonumber\\
	& = & \frac{1}{2^D} \sum_{n_i = 0,1;  p_- \in {\cal M}^{\prime}   }  \,
	\langle p + \pi n/N - p_-|\hat{A}|p + \pi n/N + p_- \rangle \,  e^{-2 i p_- q}\Pi_i \frac{1+e^{i 2 N p_i}}{2}\frac{1+e^{Ni (p^i+  \pi n /N+ p_-^i)}}{2} . \label{Buot2}
\end{eqnarray}
Here
$$
{\cal  M}^\prime = \{(m_1 \frac{2\pi}{N},...,m_D \frac{2\pi}{N}) | m_i \in \{0,1/2,1,...,N-1/2\}\}
$$
and
$$
{\cal  M}^{\prime\prime}= \{(m_1 \frac{2\pi}{N},...,m_D \frac{2\pi}{N}) | m_i \in \{0,1/4,1/2,3/4,1,...,N-1/4\}\}.
$$
The periodicity of Buot symbol receives now the form {(valid for the discrete values of arguments)}:
	$$
A_B(p,q+N n) = e^{- i 2 p N n}A_B(p,q), \quad A_B(p+\pi n,q) = e^{ 2 \pi i q n}A_B(p,q)
$$
with $n_i = 0,1$.

Let us consider the modified Buot symbol of Eq. (\ref{Buot2}). Now we impose on operator $\hat A$ the following conditions
$$
\langle q_1 | \hat{A} | q_2 \rangle = \langle q_1 +  \sum_j n^j e_j | \hat{A} | q_2 + \sum_j n^j e_j \rangle$$
where $e_j = (0,...,1,...0)$ is unity vector in the $j$ - th direction while $n_j = 0,1$.
Here $q_1, q_2 \in {\cal O}$. Now we define
$$
{\cal O}_1 = \{(m_1,...,m_D ) | m_i \in \{0,2,4,...,N-2\}\}.
$$
We also require
$$ \langle q_1 | \hat{A} | q_2 +   e_j\rangle = 0
$$
for  $q_1,q_2 \in {\cal O}_1$ and any $j$.

$\cal O$ may be divided into the two pieces ${\cal O}  = {\cal O}_1\cup {\cal O}_2$. Here
${\cal O}_2$ is the remaining piece of $\cal O$.
By $\hat{A}^1$ we denote restriction of operator $\hat{A}$ to space spanned on vectors $|q\rangle$ with $ q \in {\cal O}_1$.

Hilbert space ${\cal H}^1$ of one - particle states on the reduced momentum lattice ${\cal O}_1$ is spanned on ket vectors
$$
| q \rangle^1, q \in {\cal O}_1.
$$
Another set of basis vectors is
$$
| p \rangle^1, p \in {\cal M}^{1/2}
$$
with
$$
{\cal M}^{1/2} = \{(2\pi m_1/N,...,2 \pi m_D/N) | m_i \in \{0,1,2,...,N/2-1\}\}.
$$
We also have
$$
{\cal M}^{1/2 \prime} = \{(2\pi m_1/N,...,2 \pi m_D/N) | m_i \in \{0,1/2,1,...,N/2-1/2\}\}.
$$
Here
$$
| p \rangle^1 = \frac{1}{\sqrt{(N/2)^D}}\sum_{x\in {\cal O}_1} | x \rangle^1 e^{i p x}.
$$
We assume normalization of states
$$
^1\langle x | p\rangle^1 = \frac{1}{\sqrt{(N/2)^D}} e^{i p x},
$$
which implies completeness relations
$$
\sum_{p \in {\cal M}_1} |p \rangle^1   {}^1\langle p | = \sum_{q \in {\cal O}_1} |q \rangle^1   {}^1 \langle q |=1.
$$

One can consider inclusion of ${\cal H}^1$ into ${\cal H}$ with the following rules:
$$
| q \rangle^1 = | q \rangle.
$$
Notice that
$$
| p + e_j \pi \rangle^1 = | p \rangle^1,
$$
while
$$
| p + 2 e_j \pi \rangle = | p \rangle.
$$

Doubly modified Buot symbol of operator $\hat{A}^1$ is defined as the modified Buot symbol of operator $\hat{A}$:
\begin{eqnarray}
	A^1_{{\mathfrak B}}(p,q) &\equiv & A_{\cal B}(p,q)  =  \frac{1}{2^D} \sum_{n_i = 0,1;  p_- \in {\cal  M}^{\prime\prime} }  \,
	\langle p + \pi n/N - p_-|\hat{A}|p + \pi n/N + p_- \rangle \,  e^{-2 i p_- q}\nonumber\\&&\Pi_i \frac{1+e^{2i N (p_i+p_-^i)}}{2} \frac{1+e^{Ni (p^i+  \pi n /N+ p_-^i)}}{2} \nonumber\\ & = &
 \sum_{n_i = 0,1;  p_- \in {\cal M}^{1/2 \prime}}  \,
	^1 \langle p + \pi n/N - p_-|\hat{A}|p + \pi n/N + p_- \rangle^1 \,  e^{-2 i p_- q}\nonumber\\&&\Pi_i \frac{1 + e^{2i p^i_-   }}{2}\frac{1+e^{2i N p_i}}{2}\frac{1+e^{Ni (p^i+  \pi n /N+ p_-^i)}}{2} \frac{1+e^{2i \pi q^i}}{2}
	\label{Weyl}
\end{eqnarray}
for $q\in {\cal O}^{2 \prime}$, $p\in  {\cal M}^{\prime\prime}$. One can see, however, that this symbol is not vanishing for $q\in {\cal O}^{2 }$, $p\in  {\cal M}^{\prime}$ only.

Here we express matrix elements of operator $\hat{A}$ as follows:
\begin{eqnarray}
	\langle p_1 |\hat{A} |p_2 \rangle &=& \frac{1}{N^D} \sum_{q_1,q_2 \in {\cal O}}  \langle q_1 |\hat{A} |q_2 \rangle e^{i p_2 q_2 - i p_1 q_1}\nonumber\\
	&=& \frac{1}{N^D} \sum_{n_i = 0,1;q_1,q_2 \in {\cal O}_1 } \Big( e^{i (q_2 + n) p_2 - i (q_1 + n) p_1} \langle q_1+ n |\hat{A} |q_2+ n \rangle\Big)
	\nonumber\\
	&=& \frac{2^D}{N^D} \sum_{n_i = 0,1;q_1,q_2 \in {\cal O}_1 } e^{i p_2 q_2 - i p_1 q_1} \langle q_1 |\hat{A} |q_2 \rangle \Pi_i\frac{1 + e^{i (p^i_2  -  p^i_1)  }}{2}
	\nonumber\\
	&=& ^1\langle p_1 |\hat{A}_1 |p_2 \rangle^1 \Pi_i\frac{1 + e^{i (p^i_2  -  p^i_1)  }}{2}.
\end{eqnarray}

This definition may be easily extended to the whole range of real values of $p$ and $q$ as follows:
\begin{eqnarray}
	A_{\mathfrak B}(p,q)  & = & \sum_{p_1,p_2 \in {\cal M}^{\prime\prime}; q_1,q_2 \in  {\cal O}^{2\prime}}\frac{1}{(16 N^2)^D} {e^{2 i  p_2(q_1-q) + 2 i q_2(p-p_1)}}   A^1_{{\mathfrak B}}(p_1,q_1) .
\end{eqnarray}
With this extension we obtain the basic property of the modified Buot symbol that follows directly from the corresponding property of Buot symbol:
\begin{eqnarray}
	&&	(\hat{A}\hat{B})_{\mathfrak B}(p,q)\Big|_{p\in {\cal M}^{\prime\prime} ,q \in {\cal O}^{2\prime}}
	=  A_{\mathfrak B}(p,q)  e^{\frac{i}{2}(\overleftarrow{\partial_q}\overrightarrow{\partial_p}-\overleftarrow{\partial_p}\overrightarrow{\partial_q})}  B_{\mathfrak B}(p,q).
\end{eqnarray}

For the operator of translation to the two lattice spacings the modified Buot symbol is
\begin{eqnarray}
	T^2_j(p,q) & = &  \sum_{n_i = 0,1;  p_- \in {\cal M}^{1/2\prime\prime} }  \,
	_1 \langle p + \pi n/N - p_-|e^{2i \hat{p}_j}|p + \pi n/N + p_- \rangle_1 \,  \zg{e^{-2 i p_- q}}\nonumber\\&& \Pi_i \frac{1 + e^{2i p^i_-   }}{2}\frac{1+e^{Ni (p^i+  \pi n_i /N+ p_-^i)}}{2}\frac{1+e^{2i N p_i}}{2}\frac{1+e^{2\pi i q_i}}{2} \nonumber\\
&=&	 \sum_{n_i,m_i = 0,1;  p_- \in  {\cal M}^{1/2\prime\prime} }  \,
	\delta_{2p_-, \pi m} e^{i 2(p_j + \pi n_j/N + p_{j,-} )} \,  \zg{e^{-2 i p_- q}}\nonumber\\&&\Pi_i \frac{1 + e^{2i p^i_-   }}{2}\frac{1+e^{Ni (p^i+  \pi n_i /N+ p_-^i)}}{2}\frac{1+e^{2i N p_i}}{2}\frac{1+e^{2\pi i q_i}}{2} \nonumber\\
	&=&	 \sum_{n_i,m_i = 0,1}  \,
	 e^{i (2p_j + 2\pi n_j/N + \pi m_{j} )} \,  \zg{e^{-i \pi m q}}\nonumber\\&&\Pi_i \frac{1 + e^{i \pi m_i   }}{2}\frac{1+e^{Ni (p^i+  \pi n_i /N+ \pi m_i/2)}}{2}\frac{1+e^{2i N p_i}}{2}\frac{1+e^{2\pi i q_i}}{2}
	 \nonumber\\
	 &=&	 \sum_{n_i = 0,1}  \,
	 e^{i (2p_j + 2\pi n_j/N  )} \,  \Pi_i \frac{1+e^{Ni p^i+  i \pi n_i}}{2}\frac{1+e^{2i N p_i}}{2}\frac{1+e^{2\pi i q_i}}{2}\nonumber\\
	& = &  \Big(e^{i 2p_j}\frac{1+e^{Ni p_j}}{2}  + e^{i (2p_j + 2\pi /N  )}\frac{1-e^{Ni p_j}}{2}\Big)\Pi_i \frac{1+e^{2i N p_i}}{2}\frac{1+e^{2\pi i q_i}}{2}
	\nonumber\\
	& = & e^{i 2p_j} \Big(\frac{1+e^{i  2\pi /N  }}{2}  + e^{i N p_j}\frac{1-e^{i  2\pi /N  }}{2}\Big)\Pi_i \frac{1+e^{2i N p_i}}{2}\frac{1+e^{2\pi i q_i}}{2}.
\end{eqnarray}

We can also represent the doubly modified Buot symbol as
\begin{eqnarray}
	A^1_{{\mathfrak B}}(p,q) & = & \frac{1}{2^D} \sum_{v \in  {\cal O}^{\prime} } e^{2 i p v} \,
	\langle q-v |\hat{A} |q+v \rangle \Pi_i({1 + {e^{2i v_i \pi/N}} })\frac{1+e^{i 2\pi (q_i+v_i)}}{2}\frac{1+e^{2i N p_i}}{2}\nonumber\\ & = & \frac{1}{2^D} \sum_{n_i = 0,1;v \in  {\cal O}^{\prime} } e^{2 i p v} \,
	^1\langle q-v {-} n |\hat{A} |q+v {-}n \rangle^1 \Pi_i({1 + {e^{2i v_i \pi/N}} })\frac{1+e^{i 2\pi (q_i+v_i)}}{2}\frac{1+e^{2i N p_i}}{2}\frac{1+e^{\pi i (q_i-v_i{-}n_i)}}{2}
	\nonumber\\ & = & \frac{1}{2^D} \sum_{n_i = 0,1;v \in  {\cal O}^{} } e^{2 i p v} \,
	^1\langle q-v {-} n |\hat{A} |q+v {-}n \rangle^1 \Pi_i({1 + {e^{2i v_i \pi/N}} })\frac{1+e^{i 2\pi q_i}}{2}\frac{1+e^{2i N p_i}}{2}\frac{1+e^{\pi i (q_i-v_i{-}n_i)}}{2}. \label{Buot3}
\end{eqnarray}

\subsection{Weyl symbol on the finite lattice defined through the doubly modified Buot symbol}

The above given construction of the doubly  modified Buot symbol of an operator was  defined on the lattice with the sublattice of a lattice with the lattice spacing equal to $1$. The distance between the adjacent lattice sites of this lattice is $2$, while the number of the lattice sites within this sublattice is $(N/2)^D$. In order to build the modified Buot symbol for the lattice with unit lattice spacing and the number of lattice sites equal to $N^D$, let us first rewrite the above expression for the doubly modified Buot symbol in case of arbitrary lattice spacing $a$:
\begin{eqnarray}
	A^1_{{\mathfrak B}}(p,q)  & = &
	\sum_{n_i = 0,1;  p_- \in  {\cal M}^{1/2\prime\prime} }  \,
	^1 \langle p + \pi n/(aN) - p_-|\hat{A}|p + \pi n/(aN) + p_- \rangle^1 \,  e^{-2 i p_- q}\nonumber\\&&\Pi_i \frac{1 + e^{2i p^i_- a   }}{2}\frac{1+e^{2i N a (p_i+p_-^i)}}{2}\frac{1+e^{N a i (p^i+  \pi n_i /(aN)+ p_-^i)}}{2} \frac{1+e^{2i N  a p_i}}{2}\frac{1+e^{2\pi i q_i/a}}{2}
	\label{Weyl}
\end{eqnarray}
and
\begin{eqnarray}
	A^1_{{\mathfrak B}}(p,q) & = & \frac{1}{2^D} \sum_{n_i = 0,1;v \in  {\cal O}^{\prime} } e^{2 i p v} \,
	^1\langle q-v {-} n a |\hat{A} |q+v {-}n a \rangle^1 \nonumber\\ && \Pi_i({1 + {e^{2i v_i \pi/(aN)}} })\frac{1+e^{i 2\pi (q_i+v_i)/a}}{2}\frac{1+e^{2i N a p_i}}{2}\frac{1+e^{\pi i (q_i-v_i{-}a n_i)/a}}{2} \label{Buot3}
\end{eqnarray}
for $q\in {\cal O}^{2\prime}$, $p\in   {\cal M}^{\prime\prime}$ with
$$
{\cal M} = \{(m_1 \frac{2\pi}{Na},...,m_D \frac{2\pi}{Na}) | m_i \in \{0,1,2,...,N-1\}\}
$$
while coordinate space is
$$
{\cal O} = \{(m_1 a ,...,m_D a) | m_i \in \{0,1,2,...,N-1\}\}.
$$
We also define:
$$
{\cal M}^\prime = \{(m_1 \frac{2\pi}{Na},...,m_D \frac{2\pi}{Na}) | m_i \in \{0,1/2,1,...,N-1/2\}\}
$$
and
$$
{\cal O}_1 = \{(m_1 a,...,m_D a ) | m_i \in \{0,2,4,...,N-2\}\}.
$$
$\cal O$ may be divided into the two pieces ${\cal O}  = {\cal O}_1\cup {\cal O}_2$. Here
${\cal O}_2$ is the remaining piece of $\cal O$.
By $\hat{A}^1$ we denote restriction of operator $\hat{A}$ to space spanned on vectors $|q\rangle$ with $ q \in {\cal O}_1$.

Hilbert space ${\cal H}^1$ of one - particle states on the reduced momentum lattice ${\cal O}_1$ is spanned on ket vectors
$$
| q \rangle^1, q \in {\cal O}_1.
$$
Another set of basis vectors is
$$
| p \rangle^1, p \in {\cal M}^{1/2}
$$
with
$$
{\cal M}^{1/2} = \{(2\pi m_1/(Na),...,2 \pi m_D/(Na)) | m_i \in \{0,1,2,...,N/2-1\}\}.
$$
We also denote
$$
{\cal M}^{1/2 \prime} = \{(2\pi m_1/(Na),...,2 \pi m_D/(Na)) | m_i \in \{0,1/2,1,...,N/2-1/2\}\}
$$
and
$$
{\cal M}^{1/2 \prime \prime} = \{(2\pi m_1/(Na),...,2 \pi m_D/(Na)) | m_i \in \{0,1/4,1/2,3/4,1,...,N/2-1/4\}\}.
$$

Now we replace $a$ by $1/2$ and $N$ by $2N$. This leads us to the definition of the doubly modified Buot symbol for the lattice with $N^D$ lattice sites and unit lattice spacing:
 \begin{eqnarray}
 	A_{{ \mathfrak B}}(p,q)  & = &
 	\sum_{n_i = 0,1;  p_- \in {\cal M}^{\prime \prime} }  \,
 	 \langle p + \pi n/(N) - p_-|\hat{A}|p + \pi n/(N) + p_- \rangle \,  e^{-2 i p_- q}\nonumber\\&& \Pi_i \frac{1 + e^{i p^i_-    }}{2}\frac{1+e^{2i N (p_i+p_-^i)}}{2}\frac{1+e^{N  i (p^i+  \pi n_i /(N)+ p_-^i)}}{2}\frac{1+e^{2i N p_i}}{2}\frac{1+e^{4\pi i q_i}}{2}\nonumber\\
 	 & = &
 	 \sum_{n_i = 0,1;  p_- \in {\cal M}^{ \prime}}  \,
 	 \langle p + \pi n/(N) - p_-|\hat{A}|p + \pi n/(N) + p_- \rangle \,  e^{-2 i p_- q}\nonumber\\&& \Pi_i \frac{1 + e^{i p^i_-    }}{2}\frac{1+e^{N  i (p^i+  p_-^i) + i n_i \pi}}{2}\frac{1+e^{2i N p_i}}{2}\frac{1+e^{4\pi i q_i}}{2}
 	\label{Weyl}
 \end{eqnarray}
and
\begin{eqnarray}
	A_{{\mathfrak B}}(p,q) & = &  \sum_{n_i = 0,1;v \in  {\cal O}^{\prime\prime} } e^{2 i p v} \,
	\langle q-v {-} n/2 |\hat{A} |q+v {-}n/2 \rangle \nonumber\\ && \Pi_i\frac{{1 + {e^{2i v_i \pi/N}} }}{2}\frac{1+e^{i 4\pi (q_i+v_i)}}{2}\frac{1+e^{2i N  p_i}}{2}\frac{1+e^{4\pi i q_i}}{2}\frac{1+e^{2\pi i (q_i-v_i{-} n_i/2)}}{2} \label{Buot3}.
\end{eqnarray}
Here
$$
{\cal M}^{\prime \prime} = \{(2\pi m_1/N,...,2 \pi m_D/N) | m_i \in \{0,1/4,1/2,3/4,1,...,N-1/4\}\}
$$
$$
{\cal O}^{\prime \prime} = \{( m_1,...,m_D) | m_i \in \{0,1/4,1/2,3/4,1,...,N-1/4\}\}.
$$

This definition may be easily extended to the whole range of real values of $p$ and $q$ as follows:
\begin{eqnarray}
	A_{{ \mathfrak B}}(p,q)  & = & \sum_{p_1,p_2 \in {\cal M}^{2 \prime \prime}; q_1,q_2 \in {\cal O}^{2\prime\prime}}\frac{1}{(64 N^2)^D} {e^{2 i  p_2(q_1-q) + 2 i q_2(p-p_1)}}   A_{{ { \mathfrak B}}}(p_1,q_1) .	\label{Wcont}
\end{eqnarray}
Here
$$
{\cal M}^{2\prime \prime} = \{(2\pi m_1/(N),...,2 \pi m_D/(N)) | m_i \in \{0,1/4,1/2,3/4,1,...,2N-1/4\}\}
$$
and
$$
{\cal O}^{2 \prime\prime} = \{(m_1  ,...,m_D ) | m_i \in \{0,1/4,1/2,3/4,1,...,2N-1/4\}\}.
$$

Now we are in a position to define the lattice Weyl symbol on the refined phase space ${\cal O}^{\prime} \otimes {\cal M}^\prime $ as
\begin{eqnarray}
	A_{{ W}}(p,q)
	& = &
	\sum_{n_i = 0,1;  u \in {\cal M}^{\prime}}  \,
	\langle p + \pi n/N - u|\hat{A}|p + \pi n/N + u \rangle \,  e^{-2 i u q}\nonumber\\&& \Pi_i \frac{1 + e^{i u^i    }}{2}\frac{1+e^{N  i (p^i+  u^i) + i n_i \pi}}{2}\nonumber\\ & = &  \sum_{n_i = 0,1;v \in  {\cal O}^{\prime} } e^{2 i p v} \,
	\langle q-v {-} n/2 |\hat{A} |q+v {-}n/2 \rangle \nonumber\\ && \Pi_i\frac{{1 + {e^{2i v_i \pi/N} }}}{2}\frac{1+e^{2\pi i (q_i-v_i{-} n_i/2)}}{2}.
	\label{Weyl}
\end{eqnarray}
We can also say that
\begin{eqnarray}
	A_{{ \mathfrak B}}(p,q) &=& A_{{ W}}(p,q)\,\Pi_i\frac{1+e^{2i N  p_i}}{2}\frac{1+e^{4\pi i q_i}}{2}
\end{eqnarray}
for the argument of $A_{{ \mathfrak B}}(p,q)$ from ${\cal O}^{\prime\prime} \otimes {\cal M}^{\prime\prime} $ and
\begin{eqnarray}
	A_{{ \mathfrak B}}(p,q) &=& A_{{ W}}(p,q)\,
\end{eqnarray}
for $(q,p) \in {\cal O}^{\prime} \otimes {\cal M}^\prime $.
Similar to the definition of Buot symbol this definition is extended to the continuous values of $p$ and $q$ as
 \begin{eqnarray}
 	A_{{W}}(p,q)  & = & \sum_{p_1 \in {\cal M}^{\prime}; q_1 \in {\cal O}^{\prime}; p_2 \in {\cal M}^{\prime}; q_2 \in {\cal O}^{\prime}}\frac{1}{(4 N^2)^D} {e^{2 i  p_2(q_1-q) + 2 i q_2(p-p_1)}}   A_{{ {W}}}(p_1,q_1) 	.	\label{Wcont}
 \end{eqnarray}

Weyl symbol of translation to one lattice spacing is given by
\begin{eqnarray}
	T_j(p,q)
	& = & e^{i p_j} \Big(\frac{1+{e^{i  \pi /N} }}{2}  + e^{i N p_j}\frac{1-{e^{i  \pi /N} }}{2}\Big).
\end{eqnarray}

\subsection{Basic properties of Weyl symbol}

Basic properties of Weyl symbol follow from the properties of the modified Buot symbols.

Operator $\hat A$ defined on the ordinary lattice $\cal O$ is to be extended to the lattice ${\cal O}^{2\prime}$. Correspondingly, momentum space is extended to ${\cal M}^{2\prime}$. In the present subsection we denote the basis vectors of these extended spaces by
$$
|q{)}, \quad |p{)}, \quad q \in {\cal O}^{2\prime},p \in {\cal M}^{2\prime} .
$$
Here
$$
{(} q |p{)} = \frac{1}{\sqrt{(4N)^D}} e^{i p q}
$$
while
$$
{\langle q | p\rangle} = \frac{1}{\sqrt{N^D}} e^{i p q},\quad  q \in {\cal O}, p\in {\cal M}.
$$
For operator $\hat A$ extended to ${\cal O}^{2\prime}$ we impose constraints
$$
{(q_1} + e_j/2 | \hat{A} |q_2 + e_j/2 {)} = {(q_1}  | \hat{A} |q_2  {)}= \langle q_1 |\hat{A}_1 |q_2 \rangle \frac{1 + e^{-i (q_2  - q_1) \pi/N }}{2}, \quad {(q_1} | \hat{A} |q_2 + e_j/2 {)}=0, \quad q_1,q_2 \in {\cal O}^2
$$
(we assume here $|q+N e_j\rangle = |q\rangle$),
and
$$
{(p_1} + e_j \pi/N | \hat{A} |p_2 + e_j \pi/N {)} = {(p_1}  | \hat{A} |p_2  {)}= \langle p_1 |\hat{A}_1 |p_2 \rangle \frac{1 + e^{i (p_2  - p_1)/2}}{2}, \quad {(p_1} | \hat{A} |p_2 + e_j \pi/N {)}=0, \quad p_1,p_2 \in {\cal M}^2
$$
(we assume here $|p+2\pi e_j\rangle = |p\rangle$).

We have the following representation for $\hat{A}$:
\begin{equation}
	\hat{A} = \frac{1}{(2 \times 4N)^D}\sum_{p\in {\cal M}^{2\prime\prime}; q \in {\cal O}^{2\prime\prime}}\tilde{\Delta}(p,q) A_{{\mathfrak B}}(p,q),
\end{equation}
where
\begin{eqnarray}
	A_{\mathfrak B}(p,q) & = & \frac{1}{2^D}\sum_{v \in {\cal O}^{2\prime\prime}} e^{2 i p v} ( q-v |\hat{A} |q+v )  \Pi_i\frac{1+e^{i 4\pi (q_i+v_i)}}{2}\nonumber\\
	& = & \frac{1}{2^D}\sum_{v \in {\cal O}^{2\prime\prime}} e^{-2 i q u} ( p-u |\hat{A} |p+u )  \Pi_i\frac{1+e^{2i N (p_i+u_i)}}{2},
	\label{Buot0}
\end{eqnarray}
while
\begin{equation}
	\tilde{\Delta}(p,q) =  \sum_{u \in {\cal M}^{2\prime\prime}}e^{2 i q u}|p-u )(p+u | \Pi_i \frac{1+e^{2i N (p_i+u_i)}}{2 }.
\end{equation}

In order to derive the star property we start from expression for the Buot symbol of the product of two operators (defined for  $q\in {\cal O}^{\prime}$, $p\in {\cal M}^{\prime}$ ):
\begin{eqnarray}
	(\hat{A}\hat{B})_{\mathfrak B}(p,q)& =&\frac{1}{(2(8N)^2)^D} \sum_{p_1,p_2,u \in  {\cal M}^{2\prime\prime}; q_1,q_2 \in {\cal O}^{2\prime\prime}}(p-u|\tilde{\Delta}(p_1,q_1)\tilde{\Delta}(p_2,q_2)|p+u) \nonumber\\&& e^{-2i u q} A_{\mathfrak B}(p_1,q_1)  B_{\mathfrak B}(p_2,q_2)\Pi_i\frac{1+e^{2i N (p_i+u_i)}}{2}\nonumber\\ &= & \frac{1}{8^D}
	\sum_{n_i=0,1;p_1,p_2 \in {\cal M}^{\prime}; q_1,q_2 \in {\cal O}^{\prime}}\frac{1}{ (N)^{2D}} e^{2 i( (p_2-p)(q_1-q) + (q_2-q)(p-p_1))} \nonumber\\&&  A_{W}(p_1,q_1)  B_{W}(p_2,q_2)\Big|_{(p-p_1-p_2+\pi n) \in {\cal M}^{2}}  e^{2\pi i n(-q + q_1+q_2)}\nonumber\\
	& =& \frac{1}{8^D}\sum_{p_1,\delta p_2 \in {\cal M}^\prime; q_1,\delta q_2 \in {\cal O}^\prime}\frac{1}{ N^{2D}} e^{2 i (\delta p_2 (q_1-q) + \delta q_2 (p-p_1))} \nonumber\\&&  A_W(p_1,q_1)  e^{\delta p_2\partial_p+\delta q_2 \partial_q}  B_W(p,q)\Pi_i\Big(\frac{1 + e^{i N (\delta p^i_2 - p^i_1) }}{2} + \frac{1 + e^{i N (\delta p^i_2 - p^i_1+\pi) }}{2} e^{i 2\pi (\delta q^i_2 - q^i_1) }\Big) \nonumber\\
	& =& \frac{1}{8^D} \sum_{p_1,\delta p_2 \in  {\cal M}^\prime; q_1,\delta q_2 \in {\cal O}^\prime}\frac{1}{ N^{2D}} e^{2 i (\delta p_2 (q_1-q) + \delta q_2 (p-p_1))}\nonumber\\&&\Pi_i\Big(\frac{1 + e^{i N (\delta p^i_2 - p^i_1) }}{2} + \frac{1 + e^{i N (\delta p^i_2 - p^i_1+\pi) }}{2} e^{i 2\pi (\delta q^i_2 - q^i_1)}\Big) \nonumber\\&&  A_W(p_1,q_1)  e^{\frac{i}{2}(\overleftarrow{\partial_q}\overrightarrow{\partial_p}-\overleftarrow{\partial_p}\overrightarrow{\partial_q})}  B_W(p,q)
	\nonumber\\
	& =&   A_{W}(p,q)  e^{\frac{i}{2}(\overleftarrow{\partial_q}\overrightarrow{\partial_p}-\overleftarrow{\partial_p}\overrightarrow{\partial_q})}  B_{W}(p,q).
\end{eqnarray}
{(In the second line the exponent $e^{2 i( (p_2-p)(q_1-q) + (q_2-q)(p-p_1))}$ results from the calculation of the matrix elements of the product $\tilde{\Delta}(p_1,q_1)\tilde{\Delta}(p_2,q_2)$ taken at the discrete values $p \in {\cal M}^\prime$, $q\in {\cal O}^\prime$. It represents the kernel of the star product, and should not be confused with the kernel of the analytical continuation of Eq. (\ref{Wcont}). The latter is used in the transition to the third line, where $\delta p_2 = p_2 - p \in {\cal M}^\prime$, $\delta q_2 = q_2 - q \in {\cal O}^\prime$.)}

We come to the star identity
$$
(\hat{A}\hat{B})_{W}(p,q)|_{p\in {\cal M}^\prime, q\in {\cal O}^\prime}=A_{W}(p,q)  e^{\frac{i}{2}(\overleftarrow{\partial_q}\overrightarrow{\partial_p}-\overleftarrow{\partial_p}\overrightarrow{\partial_q})}  B_{W}(p,q).
$$
The trace identities and the other important algebraic properties of Weyl symbol follow trivially from the corresponding properties of the Buot symbol.

\subsection{Alternative derivation of star identity}
Let us give the alternative derivation of the star identity for Weyl symbol. It is assumed that $p \in {\cal M}^\prime$ while $q \in {\cal O}^\prime$. We start from the star product of Weyl symbols and come back to the Weyl symbol of the product as follows. {In the first line below the corrected analytical continuation of Eq. (\ref{Wcont}) is used in order to represent $A_W(p,q)$ through its values at the discrete points. In the second line the variables $p_2 = p+\delta p_2 \in {\cal M}^\prime$, $q_2 = q + \delta q_2 \in {\cal O}^\prime$ are introduced, which gives rise to the exponent $e^{2 i( (p_2-p)(q_1-q) + (q_2-q)(p-p_1))}$. This exponent represents the kernel of the star product, and should not be confused with the kernel of the analytical continuation.}
\begin{eqnarray}
	&&A_W(p,q)  e^{\frac{i}{2}(\overleftarrow{\partial_q}\overrightarrow{\partial_p}-\overleftarrow{\partial_p}\overrightarrow{\partial_q})}  B_W(p,q)
	 =  \sum_{p_1,\delta p_2 \in  {\cal M}^\prime; q_1,\delta q_2 \in {\cal O}^\prime}\frac{1}{ (2N)^{2D}} e^{2 i (\delta p_2 (q_1-q) + \delta q_2 (p-p_1))}A_W(p_1,q_1)  e^{\frac{i}{2}(\overleftarrow{\partial_q}\overrightarrow{\partial_p}-\overleftarrow{\partial_p}\overrightarrow{\partial_q})}  B_W(p,q)\nonumber\\&&=
	\sum_{p_1,p_2 \in {\cal M}^\prime; q_1,q_2 \in {\cal O}^\prime}\frac{1}{ (2N)^{2D}} e^{2 i( (p_2-p)(q_1-q) + (q_2-q)(p-p_1))} \nonumber\\&&  A_W(p_1,q_1)  B_W(p_2,q_2)\nonumber\\
	&&=
	\sum_{p_1,p_2 \in {\cal M}^\prime; n_1^i,n_2^i = 0,1;v_1,v_2,q_1,q_2 \in {\cal O}^\prime}\frac{1}{2^{2D}(2N)^{2D}} e^{2 i( (p_2-p)(q_1-q) + (q_2-q)(p-p_1))} \nonumber\\&&  \zg{e^{2 i p_1 v_1+2 i p_2v_2}} \langle q_1-v_1{-}n_1/2 |\hat{A} |q_1+v_1{-}n_1/2 \rangle\langle q_2-v_2{-}n_2/2 |\hat{B} |q_2+v_2{-}n_2/2 \rangle\Big|_{q_1+v_1{-}n_1/2,q_2+v_2{-}n_2/2\in {\cal O}}\nonumber\\&&\Pi_i({1 + {e^{2i v_1^i \pi/N}} })({1 + {e^{2i v_2^i \pi/N}} })  \nonumber\\
	&&=
	\sum_{n_1^i,n_2^i=0,1;v_1,v_2
		\in {\cal O}^\prime}\frac{1}{ 2^{2D}} e^{2 i p (v_1+v_2)} \nonumber\\&&  \langle q-v_1-v_2{-}n_1/2 |\hat{A} |q+v_1-v_2{-}n_1/2 \rangle\langle q +v_1-v_2{-}n_2/2 |\hat{B} |q+v_1+v_2{-}n_2/2 \rangle\Big|_{q+v_1+v_2{-}n_1/2,q+v_1-v_2{-}n_2/2\in {\cal O}} \nonumber\\&&\Pi_i({1 + {e^{2i v_1^i \pi/N}} })({1 + {e^{2i v_2^i \pi/N}} })
	\nonumber\\
	&&=
	\sum_{n^i = 0,1;v_+,v_-
		\in {\cal O}^{2\prime}}\frac{1}{ 2^{3D}} e^{2 i p v_+}    \langle q-v_+{-}n/2 |\hat{A} |q+v_-{-}n/2 \rangle\langle q +v_-{-}n/2 |\hat{B} |q+v_+{-}n/2 \rangle\Big|_{q+v_+{-}n/2,q+v_-{-}n/2,v_++v_-\in {\cal O}} \nonumber\\&&\Pi_i({1 + {e^{2i v_+^i \pi/N}} })
	\nonumber\\
	&&=
	\sum_{n^i = 0,1;v_+,v_-
		\in {\cal O}^{\prime}}\frac{1}{ 2^{D}} e^{2 i p v_+}   \langle q-v_+{-}n/2 |\hat{A} |q+v_-{-}n/2 \rangle\langle q +v_-{-}n/2 |\hat{B} |q+v_+{-}n/2 \rangle\Big|_{q+v_+{-}n/2,q+v_-{-}n/2,v_++v_-\in {\cal O}}\nonumber\\&&\Pi_i({1 + {e^{2i v_+^i \pi/N}} })
	\nonumber\\
	&&=
	\sum_{n_i=0,1;m^i=0,1;v_+,v_-
		\in {\cal O}^{\prime}}\frac{1}{ 2^{2D}} e^{2 i p v_+ + 2\pi i m (v_++v_-)} \nonumber\\&&   \langle q-v_+{-}n/2 |\hat{A} |q+v_-{-}n/2 \rangle\langle q +v_-{-}n/2 |\hat{B} |q+v_+{-}n/2 \rangle\Big|_{q+v_+{-}n/2,q+v_-{-}n/2\in {\cal O}} \nonumber\\&&\Pi_i({1 + {e^{2i v_+^i \pi/N}} }) .
	\label{D1__2}
\end{eqnarray}
One can see that for $q_i{-}n_i/2 \in Z$ both $v_+^i$ and $v_-^i$ are integer. At the same time if $q_i{-}n_i/2$ is half integer, both $v_+^i$ and $v_-^i$ are half integer. In both cases $v_+ + v_-$ is integer. As a result
\begin{eqnarray}
	&&A_W(p,q)  e^{\frac{i}{2}(\overleftarrow{\partial_q}\overrightarrow{\partial_p}-\overleftarrow{\partial_p}\overrightarrow{\partial_q})}  B_W(p,q) = \nonumber\\
	&&=
	\sum_{n_i = 0,1;v_+,v_-
		\in {\cal O}^\prime}\frac{1}{ 2^{D}} e^{2 i p v_+ }    \langle q-v_+{-}n/2 |\hat{A} |q+v_-{-}n/2 \rangle\langle q +v_-{-}n/2 |\hat{B} |q+v_+{-}n/2 \rangle\Big|_{q+v_+{-}n/2,q+v_-{-}n/2\in {\cal O}}\nonumber\\&&\Pi_i({1 + {e^{2i v_+^i \pi/N}} }) \nonumber\\
	&&=
	\zg{\sum_{n_i=0,1;\,v_+\in {\cal O}^\prime}}\frac{1}{ 2^{D}} e^{2 i p v_+ }    \langle q-v_+{-}n/2 |\hat{A} \hat{B} |q+v_+{-}n/2 \rangle\Pi_i({1 + {e^{2i v_+^i \pi/N}} }) \nonumber\\&&= (\hat{A} \hat{B})_W(p,q).
	\label{D1__2}
\end{eqnarray}

\subsection{Trace properties}

Let us express trace of an operator $\hat A$ through its Weyl symbol:
\begin{eqnarray}
&&	\frac{1}{(4N)^D} \sum_{p\in {\cal M}^\prime; q \in {\cal O}^\prime}A_W(p,q) =\nonumber\\ && =\frac{1}{2^D(4N)^D} \sum_{n^i = 0,1;p\in {\cal M}^\prime; v,q \in {\cal O}^\prime} e^{2 i p v} \langle q-v{-}n/2 |\hat{A} |q+v{-}n/2 \rangle\Big|_{q+v{-}n/2\in {\cal O}} \Pi_i({1 + {e^{2i v^i \pi/N}} })\nonumber\\ && =\frac{1}{2^D} \sum_{n^i = 0,1; q \in {\cal O}^\prime}  \langle q{-}n/2 |\hat{A} |q{-}n/2 \rangle\Big|_{q{-}n/2\in {\cal O}} \nonumber\\ && = \sum_{ r \in {\cal O}}  \langle r |\hat{A} |r \rangle = {\rm Tr}\, \hat{A}.
\end{eqnarray}
Now let us consider the trace of the product of two operators
\begin{eqnarray}
&&	\frac{1}{(4N)^D}\sum_{p\in {\cal M}^\prime;q\in {\cal O}^\prime}A_W(p,q)   B_W(p,q)
		=
	\sum_{p \in {\cal M}^\prime; n_1^i,n_2^i = 0,1;v_1,v_2,q \in {\cal O}^\prime}\frac{1}{2^{2D}(4N)^{D}}   \zg{e^{2 i p v_1+2 i p v_2}} \langle q-v_1{-}n_1/2 |\hat{A} |q+v_1{-}n_1/2 \rangle\nonumber\\&&\langle q-v_2{-}n_2/2 |\hat{B} |q+v_2{-}n_2/2 \rangle\Big|_{q+v_1{-}n_1/2,q+v_2{-}n_2/2\in {\cal O}}\Pi_i({1 + {e^{2i v_1^i \pi/N}} })({1 + {e^{2i v_2^i \pi/N}} })  \nonumber\\
	&&=
	\sum_{n_1^i,n_2^i=0,1;q,v
		\in {\cal O}^\prime}\frac{1}{ 2^{3D}} \nonumber\\&&  \langle q-v{-}n_1/2 |\hat{A} |q+v{-}n_1/2 \rangle\langle q +v{-}n_2/2 |\hat{B} |q-v{-}n_2/2 \rangle\Big|_{q+v{-}n_1/2,q-v{-}n_2/2\in {\cal O}} \nonumber\\&&\Pi_i({1 + {e^{2i v^i \pi/N}} })({1 + {e^{-2i v^i \pi/N}} })
	\nonumber\\
	&&=
	\sum_{n^i = 0,1;v_+,v_-
		\in {\cal O}^{2\prime}}\frac{1}{ 2^{3D}}     \langle v_-{-}n/2 |\hat{A} |v_+{-}n/2 \rangle\langle v_+{-}n/2 |\hat{B} |v_-{-}n/2 \rangle\Big|_{v_+{-}n/2,v_-{-}n/2,v_++v_-\in {\cal O}} \nonumber\\&&\Pi_i({1 + {\rm cos}\, (v_+^i-v_-^i) \pi/N })
	\nonumber\\
	&&=
	\sum_{n^i = 0,1;v_+,v_-
		\in {\cal O}^{\prime}}\frac{1}{ 2^{D}}   \langle v_-{-}n/2 |\hat{A} |v_+{-}n/2 \rangle\langle v_+{-}n/2 |\hat{B} |v_-{-}n/2 \rangle\Big|_{v_+{-}n/2,v_-{-}n/2,v_++v_-\in {\cal O}}
	\nonumber\\
	&&=
	\sum_{n_i=0,1;m^i=0,1;v_+,v_-
		\in {\cal O}^{\prime}}\frac{1}{ 2^{2D}} e^{ 2\pi i m (v_++v_-)} \nonumber\\&&   \langle v_-{-}n/2 |\hat{A} |v_+{-}n/2 \rangle\langle v_+{-}n/2 |\hat{B} |v_-{-}n/2 \rangle\Big|_{v_+{-}n/2,v_-{-}n/2\in {\cal O}}  \nonumber\\
	&&=
\sum_{v_+,v_-
\in {\cal O}}  \langle v_- |\hat{A} |v_+ \rangle\langle v_+ |\hat{B} |v_- \rangle = {\rm Tr}\, \hat{A}\hat{B} .
	\label{D1__2}
\end{eqnarray}

\subsection{Limit of inifinitely large lattice}

Let us consider the limit of infinitely large lattice ($N \to \infty$). Then momentum space becomes continuous.
Eq. (\ref{Weyl}) is reduced to the following expression for the Weyl symbol:
\begin{eqnarray}
	A_{{\cal W}}(p,q)  & \approx &
	\sum_{p_- \in {\cal M}^\prime }  \,
	\langle p  - p_-|\hat{A}|p  + p_- \rangle \,  {e^{-2 i p_- q}}\Pi_i \frac{1 + e^{i p^i_-    }}{2}.
	\label{WeylI}
\end{eqnarray}
Because $A_{\cal W}(p,q)$ as a function of $p$ is continuous for the operators that originate from the tight - binding models (i.e. those composed of $\hat{T}_j$ considered above), we  obtain
\begin{eqnarray}
	A_{{\cal W}}(p,q)  & \approx &
	\sum_{p_- \in {\cal M}^\prime }  \,
	\langle p  - p_-|\hat{A}|p  + p_- \rangle \,  {e^{-2 i p_- q}}\Pi_i \frac{1 + e^{i p^i_-    }}{2} \nonumber\\ & \approx & \frac{N^D}{(2\pi)^D}
	\int_{{\cal M}} d^D p_-   \,
	\langle p  - p_-|\hat{A}|p  + p_- \rangle \,  {e^{-2 i p_- q}}\Pi_i (1 + e^{i p^i_-    }).
	\label{WeylI}
\end{eqnarray}
Recall that basis vectors of Hilbert space corresponding to definite momentum are defined for finite $N$ as:
$$
| p \rangle = \frac{1}{\sqrt{(N)^D}}\sum_{q\in {\cal O}} | q \rangle e^{i p q}.
$$
Let us define the other set of basis vectors:
$$
| p \rangle\rangle \equiv \frac{1}{\sqrt{(2\pi)^D}}\sum_{q\in {\cal O}} | q \rangle e^{i p q}..
$$
In terms of these basis vectors Weyl symbol of operator is given by
\begin{eqnarray}
	A_{{\cal W}}(p,q)  & \approx &
	\int_{{\cal M}} d^D p_-   \,
	\langle\langle p  - p_-|\hat{A}|p  + p_- \rangle\rangle \,  {e^{-2 i p_- q}}\Pi_i (1 + e^{i p^i_-    }).
	\label{WeylI}
\end{eqnarray}
Extension to the real values of $p$ and $q$ in this limit obtains the form:
\begin{eqnarray}
	A_{\cal W}(p,q)  & = & \sum_{p_1,p_2 \in {\cal M}^{ \prime}; q_1,q_2 \in {\cal O}^{ \prime}}\frac{1}{(4 N^2)^D} {e^{2 i  p_2(q_1-q) + 2 i q_2(p-p_1)}}   A_{{\cal W}}(p_1,q_1) \nonumber\\ &\approx & \int_{{\cal M}} d^D p_2 d^D p_1
	\sum_{ q_1,q_2 \in {\cal O}^{\prime}}\frac{1}{( (2\pi)^2)^D} {e^{2 i  p_2(q_1-q) + 2 i q_2(p-p_1)}}   A_{{\cal W}}(p_1,q_1)
	\nonumber\\ &\approx & \int_{{\cal M}} d^D p_2
	\sum_{  q_1 \in {\cal O}^{\prime}}\frac{1}{ (2\pi)^D} {e^{2 i  p_2(q_1-q) }}   A_{{\cal  W}}(p,q_1)
	\nonumber\\ &\approx & \int_{{\cal  M}_1} d^D p_2
	\sum_{  q_1 \in {\cal O}^{\prime}}\frac{1}{ (2\pi)^D} {e^{2 i  p_2(q_1-q) }}   \int_{{\cal M}} d^D p_-   \,
	\langle\langle p  - p_-|\hat{A}|p  + p_- \rangle\rangle \,  {e^{-2 i p_- q_1}}\Pi_i (1 + e^{i p^i_-    })
	\nonumber\\ &\approx & \int_{{\cal M}} d^D p_-   \,
	\langle\langle p  - p_-|\hat{A}|p  + p_- \rangle\rangle \,  {e^{-2 i p_- q}}\Pi_i (1 + e^{i p^i_-    }).
\end{eqnarray}
One can see that our definition of Weyl symbol defined on the finite lattice is precisely reduced to the definition of \cite{FZ2020} in the limit of infinite lattice.

\section{Dynamics of systems defined on finite lattice, and Hall conductivity }
\label{Dynamics}

\subsection{Keldysh technique of field theory }

\label{SectKeldysh}


{Here we follow closely the methodology of \cite{Sugimoto2008} and \cite{BFLZZ2021}. Let us consider the inhomogeneous system defined on the $D$ - dimensional lattice $\cal O$. Time remains continuous. The complete field Hamiltonian is denoted by    $\hat\cH$.  Let the operator  $O[\psi,\bar{\psi}]$ be a functional of field operators $\hat{\psi}, \hat{\bar{\psi}}$. We suppose that $O$ is a local, i.e. at the moment of time $t$ it is a function of $\psi$ and $\bar{\psi}$  defined at the same moment. Quantum average of this operator is given by
	$$
	\langle O \rangle
	=  {\Tr} \,\Bigl(\hat{\rho}(t_i)\, e^{- i \int_{t_i}^{t} \hat\cH dt }  O[\hat{\psi},\hat{\bar{\psi}}] e^{- i \int_{t}^{t_f} \hat\cH dt } e^{ i \int_{t_i}^{t_f} \hat\cH dt }\Bigr).
	$$
	Here $t_i < t < t_f$, and  $\hat{\rho}(t_i)$ is density matrix at $t_i$. Using time ordering $T$ we rewrite it as
	$$
	\langle O \rangle =  {\Tr} \,\Bigl(T\,\Big[\hat{\rho}(t_i)\, e^{- i \int_{t_i}^{t_f} \hat\cH dt }  O[\hat{\psi},\hat{\bar{\psi}}]\Big] e^{ i \int_{t_i}^{t_f} \hat\cH dt }\Bigr).
	$$
	For the lattice model an average of quantity $O$ is given by
	$$
	\langle O \rangle
	=  \int {\cal D}\bar{\psi} {\cal D} \psi\, O[\psi,\bar{\psi}]
	\exp\left\{\ii \int_C dt \sum_x \, \bar{\psi}(t,x) \hat{Q} \psi(t,x) \right\}.
	$$
	Here $\psi$ and $\bar{\psi}$ are independent Grassmann variables, and by $x$ we understand here a $D$-dimensional lattice point. In the absence of interactions $\hat{Q}$ is given by $\hat{Q} = i \partial_t-\hat{H}$, where $\hat{H}$ is one-particle Hamiltonian.
	Integration over time $t$ is along the Keldysh contour $C$. The contour starts at the initial moment of time $t_i$, goes to the final moment $t_f$, and returns back from $t_f$ to $t_i$. The whole dynamics is concentrated between $t_i$ and $t_f$.
	
    The forward part of the contour carries the fields $\bar{\psi}_-(t,x)$ and $\psi_-(t,x)$. The fields on the  backward part are  $\bar{\psi}_+(t,x)$ and $\psi_+(t,x)$.
	
	Variables of the two parts of the Keldysh contour are independent of each other. However, there exist boundary conditions relating them to each other: $\bar{\psi}_-(t_f,x) =  \bar{\psi}_+(t_f,x)$ and $\psi_-(t_f,x)=\psi_+(t_f,x)$. The  integration measure ${\cal D} \bar{\psi} {\cal D} \psi$ contains  $\bar{\psi}_+(t_i,x)$, ${\psi}_+(t_i,x)$ and $\bar{\psi}_-(t_i,x)$, ${\psi}_-(t_i,x)$ and a weight function responsible for the initial density matrix $\hat{\rho}$:
	\begin{eqnarray}
		\langle O \rangle
		&=&
		\int \frac{{\cal D}\bar{\psi}_\pm {\cal D} \psi_\pm}{{\rm Det}\, (1+\rho)} \, O[\psi_+,\bar{\psi}_+]\nonumber\\
		&&
		\qquad	{\rm exp}\left\{\ii \int_{t_i}^{t_f} dt \sum_x \[\bar{\psi}_-(t,x) \hat{Q} \psi_-(t,x)-\bar{\psi}_+(t,x) \hat{Q} \psi_+(t,x)\]-\sum_x\, \bar{\psi}_-(t_i,x) {\rho} \psi_+(t_i,x)\right\} .\label{eq1}
	\end{eqnarray}
	Here $\rho$ (without hat) is defined as an operator in one - particle Hilbert space. Let us denote its eigenstates by $|\lambda_i \rangle$. Its matrix elements are:
 $\frac{\langle \lambda_i |\rho|\lambda_i\rangle}{1+\langle \lambda_i |\rho|\lambda_i\rangle}$ is the probability that the one - particle state $|\lambda_i\rangle$ is occupied while  $\frac{1}{1+\langle \lambda_i |\rho|\lambda_i\rangle}$ is the probability that this state is vacant.
Let us introduce Keldysh spinors
	\be
	\Psi = \left(\begin{array}{c}\psi_-\\ \psi_+ \end{array}\right),
	\label{KelPsi}
	\ee
The expression for the average of an operator $O$ receives the form 	
	\begin{eqnarray}
		\langle O \rangle
		&=& \frac{1}{{\rm Det}\, (1+\rho)}\int {\cal D}\bar{\Psi} {\cal D} \Psi \,
		O[\Psi,\bar\Psi]\,
		{\rm exp}\Bigl\{\ii \int_{t_i}^{t_f} dt \sum_x \bar{\Psi}(t,x) \hat{\bf Q} \Psi(t,x) \Bigr\} .
	\end{eqnarray}

	\rv{Again, by $x$ we understand here a $D$-dimensional vector.} Here $\bm {\hat Q}$ is in Keldysh representation
	\begin{eqnarray}
		\hat{\bf Q}
		= \left(\begin{array}{cc}Q_{--} & Q_{-+}\\ Q_{+-} & Q_{++} \end{array} \right).
		\label{KelQ}
	\end{eqnarray}
	The correct expressions for the components of this matrix may be obtained either as continuum limit of the lattice regularized expressions or using operator formalism. The result is
	\begin{eqnarray}
		Q_{++}  &=& -\Big(\ii \partial_t-\hat{H} - \ii \epsilon \frac{1-\rho}{1+\rho}\Big), \nonumber \\
		Q_{--}  &=&  \ii \partial_t-\hat{H} + \ii \epsilon \frac{1-\rho}{1+\rho}, \nonumber \\
		Q_{+-}  &=&  -2\ii \epsilon \frac{1}{1+\rho}, \nonumber \\
		Q_{-+}  &=& 2\ii  \epsilon \frac{\rho}{1+\rho}
		\label{Qnaive} .
	\end{eqnarray}
	Here $\rho$ is matrix that gives rise to initial one - particle distribution $f = \rho (1+\rho)^{-1}$. In case of the distribution depending only on energy (and, in particular for thermal distribution of non - interacting particles)  $\rho = \rho(\hat{H})$ is a function of the one - particle Hamiltonian.} The infinitely small contributions proportional to parameter $\epsilon \to 0$ symbolize the way those functions are understood as the so - called generalized functions (tempered distributions), for details see Sect. 5.1 of \cite{Kamenev2}.
	
	The Green's  function $\hat{\bf G}$ is defined as
	\begin{eqnarray}
		G_{\alpha_1 \alpha_2}(t,x|t^\prime,x^\prime)
		&=& \int \frac{{\cal D}\bar{\Psi} {\cal D} \Psi}{\ii{\rm Det}\, (1+\rho)}
		\Psi_{\alpha_1}(t,x) \bar{\Psi}_{\alpha_2}(t^\prime,x^\prime)
		\, \exp\left\{\ii \int_{t_i}^{t_f} dt \sum_x\, \bar{\Psi}(t,x) \hat{\bf Q} \Psi(t,x) \right\}.\label{G1}
	\end{eqnarray}
		Here index $\alpha$ corresponds to components of Keldysh spinor  \Ref{KelPsi}. The Green function obeys equation
	$$
	\hat{\bf Q}\hat{\bf G}=1 .
	$$
	For the components of $\hat{\bf G}$ we have the following relation
	\be
	G_{--}+G_{++}-G_{-+}-G_{+-}=0
	\label{G-rel}
	\ee
	while 	
	\be	
	Q_{--}+Q_{++}+Q_{-+}+Q_{+-}=0 .
	\label{Q-rel}
	\ee
	
	Sometimes the new representation of Keldysh spinors is used that is related to the spinors defined above as follows	
	$$
	\begin{pmatrix}\psi_1 \\ \psi_2 \end{pmatrix}=\frac{1}{\sqrt{2}}\begin{pmatrix}1 & 1 \\ 1 & -1 \end{pmatrix}\begin{pmatrix}\psi_- \\ \psi_+ \end{pmatrix}
	,\quad
	\begin{pmatrix}\bar{\psi}_1 ,& \bar{\psi}_2 \end{pmatrix}=\frac{1}{\sqrt{2}}\begin{pmatrix}\bar{\psi}_-, & \bar{\psi}_+ \end{pmatrix}\begin{pmatrix}1 & 1 \\ -1 & 1 \end{pmatrix}.
	$$
	Green function in the new representation receives the triangle form
	\begin{eqnarray}
		\hat{\bf G}^{(K)}&=&-i\langle \begin{pmatrix}\psi_1 \\ \psi_2 \end{pmatrix}\otimes\begin{pmatrix}\bar{\psi}_1, & \bar{\psi}_2 \end{pmatrix}\rangle \\\nonumber&=&\frac{1}{2}\begin{pmatrix}1 & 1 \\ 1 & -1 \end{pmatrix}\begin{pmatrix}G^{--} & G^{-+} \\ G^{+-} & G^{++} \end{pmatrix}\begin{pmatrix}1 & 1 \\ -1 & 1 \end{pmatrix}\\\nonumber
		&=&\begin{pmatrix}
			G^\rR &G^\rK \\0&G^\rA
		\end{pmatrix}.
		\label{GK}
	\end{eqnarray}
	Here the Keldysh, Advanced and Retarded Green  functions are introduced:
	\bes
	G^\rK  & =G^{-+}+G^{+-}=G^{--}+G^{++},\\
	G^\rA  & =G^{--}-G^{+-}=G^{-+}-G^{++},\\
	G^\rR  & =G^{--}-G^{-+}=G^{+-}-G^{++}.
\end{eqsplit}

In our paper we will use yet another representation
\bes
\hat{\bf G}^{(<)}&=\begin{pmatrix}
1&1\\0&1
\end{pmatrix}\begin{pmatrix}
G^\rR &G^\rK \\0&G^\rA
\end{pmatrix}\begin{pmatrix}
1&-1\\0&1
\end{pmatrix}
\\
&=	\begin{pmatrix}
G^\rR &2G^<\\0&G^\rA
\end{pmatrix} .
\label{G<}	
\end{eqsplit}
It is related to the Green function defined by Eq. (\ref{G1}) as follows
\be
\hat{\bf G}^{(<)} = U \hat{\bf G} V,
\ee
where
$$
U=\frac{1}{\sqrt{2}}\begin{pmatrix}
1&1\\0&1
\end{pmatrix}\begin{pmatrix}1 & 1 \\ 1 & -1 \end{pmatrix}=\frac{1}{\sqrt{2}}\begin{pmatrix}2 & 0\\1 &-1 \end{pmatrix}
$$
and
$$
V=\frac{1}{\sqrt{2}}\begin{pmatrix}
1&1\\-1&1
\end{pmatrix}\begin{pmatrix}1 & -1 \\ 0 & 1 \end{pmatrix}=\frac{1}{\sqrt{2}}\begin{pmatrix}1 & 0\\-1& 2 \end{pmatrix}.
$$
In addition, we have
\begin{eqnarray}
\hat{\bf Q}^{(<)}&=&V^{-1}\hat{\bf Q} U^{-1}\\\nonumber&=&\frac{1}{2}\begin{pmatrix}2 & 0\\1& 1 \end{pmatrix}\begin{pmatrix}Q^{--} & Q^{-+} \\ Q^{+-} & Q^{++} \end{pmatrix}\begin{pmatrix}1 & 0\\1& -2 \end{pmatrix}\\\nonumber&=&
\begin{pmatrix}Q^{--}+Q^{-+} & -2Q^{-+} \\ \frac{Q^{--}+Q^{+-}+Q^{-+}+Q^{++}}{2} & -Q^{-+}-Q^{++} \end{pmatrix}\\\nonumber&=&\begin{pmatrix}Q^\rR  & 2Q^< \\ 0 & Q^\rA  \end{pmatrix},
\end{eqnarray}
Here we denote
\be
Q^\rR =Q^{--}+Q^{-+}, \qquad
Q^\rA =-Q^{-+}-Q^{++}, \qquad
Q^<=-Q^{-+}.
\label{Qar_def}
\ee
As a result
\begin{equation}
G^\rA  = (Q^\rA )^{-1}, \qquad G^\rR  = (Q^\rA )^{-1}, \qquad G^< =-G^\rR  Q^< G^\rA .
\label{Gar_def}
\end{equation}
with
\bes
G^\rR
&= (\ii \partial_t-\hat{H}e^{+ \epsilon \partial_t})^{-1}
= (\ii \partial_t-\hat{H}+\ii \epsilon )^{-1},
\\
G^\rA  &= ( \ii \partial_t-\hat{H}e^{- \epsilon \partial_t})^{-1}
= ( \ii \partial_t-\hat{H}-\ii \epsilon )^{-1},
\\
G^< &=(G^\rA -G^\rR ) \frac{\rho}{\rho+1}.
\label{Gar_expl}
\end{eqsplit}
The elements of $\hat{\bf Q}^<$ (which is inverse to $\hat{\bf G}^<$) are:
\bes
Q^{<}&=(Q^\rA -Q^\rR )\frac{\rho}{\rho+1} = -2\ii\epsilon \frac{\rho}{\rho+1},
\\
Q^\rR  &= \ii \partial_t-\hat{H}+\ii \epsilon ,	
\\
Q^\rA  &=  \ii \partial_t-\hat{H}-\ii \epsilon .
\label{Qar_expl}
\end{eqsplit}
For more details on the basics of Keldysh technique briefly reviewed above the reader is advised to consult  \cite{Kamenev,Kamenev2}.

\subsection{ Keldysh technique in terms of Weyl symbol, and conductivity}

Basic notions of Wigner - Weyl calculus may be found, for example, in \cite{ZW2019,Sugimoto}. Here we adopt them to the models defined on {\it finite} lattice. In the following the $D+1$ dimensional vectors (with space and time components) are denoted by large Latin letters. We denote matrix element of an operator $\hat{A}$ by  $A(X_1,X_2) = \langle X_1 | \hat{A} | X_2 \rangle $. Since we deal with the lattice models the space components of $D+1$ - vectors are discrete while the time components are continuous. We then define the Weyl symbol of an operator $\hat A$ as the mixture of Weyl symbol (with respect to discrete space components) defined above,  and the standard Wigner transformation with respect to the time component:
\begin{eqnarray}
A_W(X|P)&=&2\int d Y^0\, \sum_{n_i = 0,1;\vec{Y} \in  {\cal O}^\prime} e^{2 \ii Y^\mu P_\mu }  A(X+Y{-}n/2,X-Y{-}n/2) \Pi_{i=1...D}\frac{{1 + {e^{-2i Y_i \pi/N}} }}{2}\frac{1+e^{2\pi i (X_i+Y_i{-} n_i/2)}}{2}, \nonumber \\ && \quad \mu =0,1,...,D. \label{WignerTr}
\end{eqnarray}
$D+1$ momentum is denoted by $P^\mu=(P^0,p)$, and $P_\mu = (P^0,-p)$. Here  $p$ is spatial momentum with $D$ components.
Below Weyl symbol of Keldysh Green function $\hat{\bf G}$ is denoted by $\hat{G}$, while Weyl symbol of Keldysh $\hat{\bf Q}$ is $\hat{Q}$. We omit the subscript $W$ for brevity in this section.

Weyl symbols $\hat{G}$ and $\hat{Q}$ obey Groenewold equation
\begin{equation}
\hat{Q} * \hat{G} = 1_W.
\end{equation}
 Here the Moyal product $*$ is defined as
\begin{equation}
\left(A* B\right)(X|P) = A(X|P)\,e^{\rv{-}\ii(\overleftarrow{\partial}_{X^{\mu}}\overrightarrow{\partial}_{P_{\mu}}-\overleftarrow{\partial}_{P_{\mu}}\overrightarrow{\partial}_{X^{\mu}})/2}B(X|P).
\end{equation}
In the present paper we consider the situation when electromagnetic potential $A$ corresponds to constant components of field strength ${\cal F}^{\mu\nu}$. Moreover, the gauge is chosen, in which the spatial part of potential is proportional to time but does not depend on spatial coordinates. Expansion in powers of
${\cal F}^{\mu \nu}$ will be used up to the leading order, proportional to electric  field. Introduction of such a form of external gauge potential  results in Peierls substitution $P \to \pi = P- A$. Here $\pi^\mu$ is $D+1$ - dimensional vector similar to $P^\mu$. When index is lowered, its spatial components change the sign. The Moyal product may be decomposed as
\begin{equation}
* =  \star~ e^{\rv{-}\ii  \mathcal{F}^{\mu\nu}\overleftarrow{\partial}_{\pi^{\mu}}\overrightarrow{\partial}_{\pi^{\nu}}/2}.
\end{equation}
with
\begin{equation}
\left(A\star B\right)(X|\pi) = A(X|\pi)\,e^{\rv{-\ii(\overleftarrow{\partial}_{X^{\mu}}\overrightarrow{\partial}_{\pi_{\mu}}-\overleftarrow{\partial}_{\pi_{\mu}}\overrightarrow{\partial}_{X^{\mu}})/2}}B(X|\pi).
\end{equation}
This expression remains valid specifically for the case of the given external field $A$ that does not depend on spatial coordinates, but depends on time giving rise to external electric field. The case of the external field depending on spatial coordinates would be more involved due to the presence of spatial lattice.

Next, we use expansion of $\hat{Q}$ and $\hat{G}$ in powers of  $\mathcal{F}^{\mu\nu}$ and keep the terms up to the linear
\begin{equation}
\hat{Q} = \hat{Q}^{(0)}  +\frac{1}{2}\mathcal{F}^{\mu\nu}\hat{Q}_{\mu\nu}^{(1)},\quad
\hat{G} = \hat{G}^{(0)}  +\frac{1}{2}\mathcal{F}^{\mu\nu}\hat{G}_{\mu\nu}^{(1)}.\label{QGK}
\end{equation}
Besides, below we omit for simplicity the superscript $^{(0)}$ of the zeroth order contribution to both $G$ and $Q$. For the case of Hamiltonian that does not depend on time and initial distribution $f(\pi_0)$ depending on energy only we have
\bes
G^\rR
&=(\pi_0-\hat{H}(\vec{\pi},x)+\ii \epsilon )^{-1},
\\
G^\rA  &= ( \pi_0-\hat{H}(\vec{\pi},x)-\ii \epsilon )^{-1},
\\
G^< &=(G^\rA -G^\rR ) f(\pi_0) = 2\pi i \delta(\pi_0-\hat{H}(\vec{\pi}))f(\pi_0).
\label{Gar_expl}
\end{eqsplit}
Matrix $\hat{\bf Q}^<$ is inverse to $\hat{\bf G}^<$ (with respect to the Moyal product):
\bes
Q^{<}&=(Q^\rA -Q^\rR )f(\pi_0) = -2\ii\epsilon f(\pi_0),
\\
Q^\rR  &= \pi_0-\hat{H}(\vec{\pi},x)+\ii \epsilon ,	
\\
Q^\rA  &=  \pi_0-\hat{H}(\vec{\pi},x)-\ii \epsilon .
\end{eqsplit}
The Groenewold equation can be written as
\begin{equation}
\left(\hat{Q}  +\frac{1}{2}\mathcal{F}^{\mu\nu}\hat{Q}_{\mu\nu}^{(1)}\right)\star~ e^{\rv{-}i  \mathcal{F}^{\mu\nu}\overleftarrow{\partial}_{\pi^{\mu}}\overrightarrow{\partial}_{\pi^{\nu}}/2}\left(\hat{G}  +\frac{1}{2}\mathcal{F}^{\mu\nu}\hat{G}_{\mu\nu}^{(1)}\right) = 1_W.
\label{Groe-F}
\end{equation}
In the zeroth order in $\cal F$ the Groenewold equation is reduced to $\hat{Q} \star \hat{G}  = 1_W$, while the first order reads
$\hat{Q} \star \hat{G}^{(1)}+\hat{Q}^{(1)}\star\hat{G} \rv{-} \ii \hat{Q} \star \overleftarrow{\partial}_{\pi^{\mu}}\overrightarrow{\partial}_{\pi^{\nu}} \hat{G}  = 0$. We come to
\begin{equation}
\hat{G}_{\mu\nu}^{(1)} =-\hat{G} \star  \hat{Q}_{\mu\nu}^{(1)}\star \hat{G}   \rv{-}   \ii\left(\hat{G} \star \partial_{\pi^{\mu}}\hat{Q}  \star\hat{G} \star \partial_{\pi^{\nu}}\hat{Q} \star \hat{G} -(\mu\leftrightarrow \nu)\right)/{2}
.
\label{QGK1}
\end{equation}

The derivation presented here is similar to the one of \cite{Sugimoto}. However, it differs essentially because we consider the model defined on the finite lattice. In the case of a uniform system Weyl symbol of operator $\hat{Q}$ does not depend on $x\in {\cal O}^\prime$. It may depend on $P^0$ and $X^0$ and on $p \in {\cal M}^\prime$. Electric current density on the lattice may be calculated as
$$
\hat{j^i}=-\hat{\bar{\psi}} \frac{\partial \hat{Q}}{\partial p_i} \hat{\psi},\quad
i=1,2,\ldots D.
$$
Spatial components of momentum are $p^i=p_i = P^i = - P_i$.
\begin{eqnarray}
	\langle j^i(t,x) \rangle
	&=&-\frac{\ii}{2}{\tr} \[\hat{\bf G} \hat{{\bf v}}^i\].
\end{eqnarray}
Velocity operator is given by
$$
\hat{\bf v}^i = \partial_{p_i}\begin{pmatrix}-Q^{--} & 0 \\ 0 & Q^{++} \end{pmatrix}.
$$
{In these expressions the derivative of the Weyl symbol with respect to momentum replaces the derivative with respect to the external gauge potential. The two derivatives coincide up to the artificial terms that disappear in the limit of infinite lattice. Therefore, here and in the expressions given below (including the one for $\cal N$) it is assumed that inside the Weyl symbol of $\hat{\bf Q}$ the limit $N \to \infty$ is taken before the differentiation with respect to momentum.}

In the presence of inhomogeneity the given expression for electric current density is already not valid. However, we can still calculate the response of the partition function to constant in space variation of electromagnetic potential. Such a variation gives expression for the electric current averaged over the whole lattice as
\begin{eqnarray}
	\langle J^i(t) \rangle
	&=&-\frac{\ii}{2} \int \frac{dP^0}{2\pi}  \frac{1}{(2N)^{2D}} \sum_{p \in   {\cal M}^\prime, x\in {\cal O}^\prime} {\tr}\,  {\bf G}(X|P) \partial_i \begin{pmatrix}-Q^{--}(X|P) & 0 \\ 0 & Q^{++}(X|P) \end{pmatrix}.
\end{eqnarray}
In the last expression we already deal with the Weyl symbols of operators.

The following quantity defined in phase space
$$
\hat{\bf v}^i = \partial_{p_i}\begin{pmatrix}-Q^{--}(P|X) & 0 \\ 0 & Q^{++}(P|X) \end{pmatrix}.
$$
may be considered as Weyl symbol of velocity operator. Therefore, we can write
\begin{eqnarray}
	\langle J^i(t) \rangle
	&=&-\frac{\ii}{2 N^D} \int \frac{dP^0}{2\pi} {\Tr} \[\hat{\bf G} \hat{{\bf v}}^i\].
\end{eqnarray}
Here $\Tr$ contains trace over the Keldysh indexes as well as trace over spatial discrete coordinates.

Let us express average current $J$  through the Keldysh Green function written in triangle representation of Eq. (\ref{G<}).
\begin{eqnarray}
\hat{\bf v}_i^{(<)}&=&\partial_{p_i}\frac{1}{2}\begin{pmatrix}2 & 0\\1& 1 \end{pmatrix}\begin{pmatrix}-Q^{--} & 0 \\ 0 & Q^{++} \end{pmatrix}\begin{pmatrix}1 & 0\\1& -2 \end{pmatrix}\\\nonumber
&=&\partial_{p_i}\begin{pmatrix} -Q^{--} & 0 \\ \frac{-Q^{--}+Q^{++}}{2}  & -Q^{++} \end{pmatrix}.
\end{eqnarray}
Using Eq. \Ref{Qar_def} ($Q^{--}=Q^\rR +Q^<$, $Q^{-+}=-Q^<$, $Q^{+-}=-Q^\rR +Q^\rA -Q^<$, and $Q^{++}=Q^<-Q^\rA $) we represent current density as
\begin{eqnarray}
\langle J^i \rangle
&=&- \frac{\ii}{2 N^D}  \int \frac{dP^0}{2\pi}  \Tr\left[\hat{\bf G}\hat{\bf v}^i\right]
=-\frac{\ii}{2 N^D} \int \frac{dP^0}{2\pi} \Tr\left[\begin{pmatrix}
G^\rR &2G^<\\0&G^\rA
\end{pmatrix}\partial_{p_i}\begin{pmatrix}- Q^\rR -Q^< & 0 \\ -\frac{Q^\rR +Q^\rA }{2}  & -Q^<+Q^\rA  \end{pmatrix}\right]\\\nonumber
&=&\frac{\ii}{2 N^D} \int \frac{dP^0}{2\pi} \Tr \left(G^\rR \partial_{p_i} Q^\rR -G^\rA \partial_{p_i}Q^\rA \right)
+\frac{\ii}{2 N^D} \int \frac{dP^0}{2\pi} \Tr \left(G^\rR \partial_{p_i} Q^<+G^<\partial_{p_i} Q^\rA \right)\nonumber\\&&
+\frac{\ii}{2 N^D} \int \frac{dP^0}{2\pi} \Tr \left(G^\rA \partial_{p_i}Q^< +G^<\partial_{p_i} Q^\rR \right).
\end{eqnarray}
The second term here is expressed through $\frac{\ii}{2}\Tr\left({G}\partial_{p_i} {Q}\right)^<$.
At the same time the third term is its complex conjugate. We obtain
\be
\langle J^i \rangle
=\frac{\ii}{2 N^D} \int \frac{dP^0}{2\pi}  \Tr\left(\hat{\bf G}\partial_{p_i}\hat {\bf Q}\right)^\rR
+ \frac{\ii}{2 N^D} \int \frac{dP^0}{2\pi}  \Tr\left(\hat{\bf G}\partial_{p_i}\hat {\bf Q}\right)^<+{\rm c.c.}
\label{<j>}
\ee
We represent electric current as
\bes
\langle J^i(t) \rangle
& 	= \rv{-} \frac{\ii }{2}\frac{1}{(2N)^{2D}} \int \frac{dP^0}{2\pi}  \sum_{p \in   {\cal M}^\prime, x\in {\cal O}^\prime}
\tr\left(\hat{G} (\partial_{\pi_{i}}\hat{Q})\right)^{\rR}
\rv{-} \frac{\ii }{2}\frac{1}{(2N)^{2D}} \int \frac{dP^0}{2\pi}  \sum_{p \in   {\cal M}^\prime, x\in {\cal O}^\prime}
\tr\left(\hat{G} (\partial_{\pi_{i}}\hat{Q})\right)^{\rA}	
\\
&\qquad		
\rv{-}\frac{\ii }{2}\frac{1}{(2N)^{2D}} \int \frac{dP^0}{2\pi}  \sum_{p \in   {\cal M}^\prime, x\in {\cal O}^\prime}
\tr\left(\hat{G} (\partial_{\pi_{i}}\hat{Q})\right)^{<}
\rv{-}\frac{\ii }{2}\frac{1}{(2N)^{2D}} \int \frac{dP^0}{2\pi}  \sum_{p \in   {\cal M}^\prime, x\in {\cal O}^\prime}
\tr\left((\partial_{\pi_{i}}\hat{Q}) \hat{G}\right)^{<} .
\end{eqsplit}
Due to the term proportional to $\pm \ii \epsilon$ the poles of $G^{\rR}$  ($G^{\rA}$) are shifted out of the real axis of $\omega$. The integration contour  may be closed at infinity. For that we need to use lattice regularization of time. The lattice regularization adds the  factor that suppresses expressions inside the integral over (complex - valued) $\omega$ for $|\omega| \to \infty$. This results in vanishing of the sum of the first two terms in the above expression.
We obtain
\be
J^i(t) = \rv{-}\frac{\ii }{2}\frac{1}{(2N)^{2D}} \int \frac{dP^0}{2\pi}  \sum_{p \in   {\cal M}^\prime, x\in {\cal O}^\prime}
\tr\left(\hat{G} (\partial_{\pi_{i}}\hat{Q})\right)^{<}
\rv{-}\frac{\ii }{2}\frac{1}{(2N)^{2D}} \int \frac{dP^0}{2\pi}  \sum_{p \in   {\cal M}^\prime, x\in {\cal O}^\prime}
\tr\left((\partial_{\pi_{i}}\hat{Q}) \hat{G}\right)^{<} .
\label{J Wigner}
\ee
Applying Eqs. (\ref{QGK})-(\ref{QGK1}) we calculate the contribution to electric current proportional to external field strength $\mathcal{F}^{\mu\nu}$:
\begin{eqnarray}
{J}^i
&=&    -\frac{1}{4}\frac{1}{(2N)^{2D}} \int \frac{dP^0}{2\pi}  \sum_{p \in   {\cal M}^\prime, x\in {\cal O}^\prime} \tr\Bigl(\hat{G} \star \partial_{\pi^{\mu}}\hat{Q}  \star\hat{G} \star \partial_{\pi^{\nu}}\hat{Q} \star \hat{G}  \partial_{\pi_{i}}\hat{Q} \Bigr)^{<}\mathcal{F}^{\mu\nu}\nonumber\\
&&
-\frac{1}{4}\frac{1}{(2N)^{2D}} \int \frac{dP^0}{2\pi}  \sum_{p \in   {\cal M}^\prime, x\in {\cal O}^\prime} \tr\Bigl(\partial_{\pi_{i}}\hat{Q}  \hat{G} \star \partial_{\pi^{\mu}}\hat{Q}  \star\hat{G} \star \partial_{\pi^{\nu}}\hat{Q} \star \hat{G}  \Bigr)^{<}\mathcal{F}^{\mu\nu}.
\end{eqnarray}

Assuming that $\cal F$ includes only electric field, we represent this expression in two-dimensional systems as:
$$
{J}^i = \sigma^{ij}  \mathcal{F}_{0j} ,
$$
where the conductivity tensor $\sigma^{ij}$ may be given as follows:
\begin{equation}
	\sigma^{ij} =  {\frac{1}{4}} \frac{1}{(2N)^{2D}} \int \frac{dP^0}{2\pi}  \sum_{p \in   {\cal M}^\prime, x\in {\cal O}^\prime} \tr\left(\partial_{\pi_{i}}\hat{Q}_W  \left[\hat{G}_W \star \partial_{\rv{\pi_{[0}}}\hat{Q}_W  \star \partial_{\rv{\pi_{j]}}}\hat{G}_W  \right]\right)^< +{\rm c.c.}\label{MAIN}
\end{equation}
In this expression we restore subscript index $W$ for the Weyl symbols. Here $(...)_{[0} (...)_{ j]} =(...)_{0} (...)_{ j} -(...)_{j} (...)_{ 0}  $ means anti-symmetrization.
Conductivity may be expressed as a sum of symmetric and anti-symmetric parts
$\sigma^{ij} = \sigma^{ij}_{H} + \sigma^{ij}_S$. Here asymmetric part  $\sigma^{ij}_H = (\sigma^{ij}-\sigma^{ji})/2$ is the Hall  conductivity while $\sigma^{ij}_S = (\sigma^{ij}+\sigma^{ji})/2$ is the conventional conductivity.

\subsection{Equilibrium limit of Hall conductivity}
\label{SectEquilibrium}

One of the basic properties of Weyl symbol is that the star may be inserted between the two Weyl symbols standing under the trace if the sum over the whole phase space is added \Ref{MAIN}. Our lattice version of Weyl symbol also obeys this property. Therefore, after averaging the conductivity over the whole volume of the system and over the overall time of the process, we obtain
\be
\bar \sigma^{ij}
=  \rv{-}{\frac{1}{4}} \frac{1}{(2N)^{2D}} \int \frac{dP^0 dX^0}{2\pi (t_f-t_i)}  \sum_{p \in   {\cal M}^\prime, x\in {\cal O}^\prime}
\tr\left(\partial_{\pi_{i}}\hat{Q}_W \star \hat{G}_W \star \partial_{\pi_{[0}}\hat{Q}_W  \star \hat{G}_W\star \partial_{\pi_{j]}}\hat{Q}_W\star\hat{G}_W\right)^< +{\rm c.c.}
\ee
The next step is to assume that $\hat Q$ entering the above expression does not depend on time. Recall that the external electromagnetic potential depends on time giving rise to electric field. But after we performed expansion in powers of electric field, this time dependence disappears from our expressions. For the initial thermal distribution we are able to represent the integral over the frequency as a sum over Matsubara frequencies. By $\Pi$ we denote  Euclidean $D+1$ - momentum, i.e. $\Pi^{D+1} = \omega$ is Matsubara frequency, while $\Pi^i = \pi^i$ for $i=1,...,D$. Notice that $\partial_{\pi^0} = -\ii \partial_{\Pi^{D+1}}$. Substituting $i\omega$ instead of $\pi^0$ we obtain the Matsubara Green function $G^M$ instead of the advanced or retarded Green function.

More specifically, for the system with the one - particle Hamiltonian $\hat{H}$ one can define the real time Green function as
	$$
	G(x_1,x_2,\omega) \equiv \langle x_1|(\omega - \hat{H})^{-1}|x_2\rangle
	$$
This Green function gives rise to Advanced, Retarded or time ordered Green function when the integration contour in plane of complex $\omega$ is shifted in a specific way. Namely,
the time ordered Green's  function (Feynman propagator) is
\begin{equation}
	G^\rT ( x, x^{\prime}, \omega)
	= \lim\limits_{\eta\rightarrow 0} G(x, x^{\prime}, \omega+\ii\eta \, {\rm sign} \,\omega).
\end{equation}
The retarded Green's  function is defined as
\begin{equation}
	G^\rR ( x, x^{\prime}, \omega) = \lim\limits_{\eta\rightarrow 0} G(x, x^{\prime}, \omega+\ii\eta),
\end{equation}
The advanced Green's function is given by
\begin{equation}
	G^\rA ( x, x^{\prime}, \omega) = \lim\limits_{\eta\rightarrow 0} G(x, x^{\prime}, \omega-\ii\eta).
\end{equation}
The Matsubara Green's function $G^\rM $ is then defined as
\be
G^\rM ( x, x^{\prime},\omega_n) = G( x, x^{\prime},\ii\omega_n),
\label{Mats}
\ee
We may write it in terms of imaginary time $\tau$:
\begin{equation}
	G^\rM ( x, x^{\prime},\tau)
	= \frac{1}{\beta}\sum\limits_{n=-\infty}^{\infty}e^{-\ii\omega_n \tau}G( x, x^{\prime},\ii\omega_n).
\end{equation}
Here $\omega_n = (2n+1)\pi/{\beta}$ is the Matsubara frequency while $\beta= 1/T$ is the inverse of temperature.

 The conductivity of the $2$ - dimensional system (averaged over the system area) is given by
$$
\bar{\sigma}^{ij}  = \frac{{\cal N}}{2\pi}\epsilon^{ij},
$$
where
\begin{eqnarray}
	{\cal N} &=&2\pi  \, T \, \frac{1}{3! \, }\epsilon^{\mu\nu\rho} \frac{1}{(2N)^{2D}}   \sum_{p \in   {\cal M}^\prime, x\in {\cal O}^\prime} \sum_{\omega_n = 2\pi T(n+1/2)}
	\tr\left(\partial_{\rv{\Pi^{\mu}}}\hat{Q}_W ^\rM  \star\hat{G}_W ^\rM \star \partial_{\rv{\Pi^{\nu}}}\hat{Q}_W ^\rM  \star\hat{G}_W ^\rM \star \partial_{\rv{\Pi^{\rho}}}\hat{Q}_W ^\rM \star \hat{G}_W ^\rM \right).
\end{eqnarray}
Here  $\epsilon^{ij}$ and $\epsilon^{\mu\nu\rho}$  are standard antisymmetric tensors. $Q_W^M$ is inverse to Matsubara Green function $G_W^M$:
$$
Q_W^M \star G_W^M = 1_W  = 1
$$

In case of small temperatures the sum over Matsubara frequencies is reduced to an integral, which gives
\begin{eqnarray}
	{\cal N} &=& \frac{1}{3!\,}\epsilon^{\mu\nu\rho} \frac{1}{(2N)^{2D}} \int {d\Pi^3}  \sum_{p \in   {\cal M}^\prime, x\in {\cal O}^\prime}
	\tr\left(\partial_{\rv{\Pi^{\mu}}}\hat{Q}_W ^\rM  \star\hat{G}_W ^\rM \star \partial_{\rv{\Pi^{\nu}}}\hat{Q}_W ^\rM  \star\hat{G}_W ^\rM \star \partial_{\rv{\Pi^{\rho}}}\hat{Q}_W ^\rM \star \hat{G}_W ^\rM \right).
	\label{NEQ}
\end{eqnarray}
(Here $D=2$.)
At any given finite value of $N$ the given expression is not robust to smooth variations of the system. However, when we increase the lattice size in such a way that it becomes much larger than any correlation length existing in the given system, then effectively the sum over momenta in the above expression may be replaced by an integral. In this case $N$ is very large but still finite. Eq. (\ref{NEQ}) remains well defined and finite and becomes topological invariant due to the presence of an integral over $p$.  Notice, that the sum over $x$ is important for the topological invariance of this quantity. {As explained above, inside the Weyl symbols entering Eq. (\ref{NEQ}) the limit of infinite $N$ is to be taken before the differentiation with respect to momenta.}

\section{Conclusions}

\label{Concl}

In the present paper we construct the Wigner - Weyl calculus for the tight - binding models defined on the finite lattices. Our starting point was the formalism proposed long time ago by Felix Buot. On the intuitive level his construction gives the hint on how to construct the formalism that allows to describe effectively dynamics of those models. However, the very definition of the symbol of operator proposed by Buot fails to reproduce relations obeyed by the Weyl symbol of operator defined in continuous theory. We, therefore, correct the Buot's definition and define the symbol of operator that is defined not only for the momenta and coordinates that belong to the original discrete phase space ${\cal O} \otimes {\cal M}$. Our first symbol of operator is defined on the refined discrete momentum space  ${\cal O}^\prime \otimes {\cal M}^\prime$ that contains $2^{2D}$ larger number of points. Moreover, the explicit formula for the extension of this definition to the continuous values of momenta and coordinates is given. The resulting symbol of operator obeys some of the basic properties that repeat those of the continuous Weyl symbol. We feel this appropriate to call this symbol the {\it Buot symbol} of operator defined on a finite lattice.

Although the Buot symbol of operator obeys several beautiful properties, it has certain disadvantages. Namely, for the simplest tight - binding models the Buot symbol of an elementary translation to one lattice spacing appears to oscillate fast both as a function of momenta and as a function of coordinates. The limit of this quantity is not regular, when the lattice size tends to infinity. The same refers to Buot symbol of unity operator. In principle, it might be possible to build the Wigner - Weyl field theory based on the Buot symbol. Then all quantities may be expressed through the Buot symbols of operators using the Moyal products. However, the limit of large lattice volume of the obtained expressions cannot be taken. For example, we cannot demonstrate in a transparent way how the Hall conductivity tends to a topological quantity in the limit of infinitely large lattice for the thermal equilibrium at small temperatures.

One of the purposes of the present construction (as mentioned in the Introduction) is to build the finite volume regularization of the lattice Wigner - Weyl calculus defined on the inifinite lattice proposed in \cite{FZ2020}. The latter construction  admits the topological expression for the Hall conductivity. However, the averaging over the lattice volume remains badly defined. We need the definition of Weyl symbol of operator for the finite lattice that tends in a regular way to the one of \cite{FZ2020} in the limit of infinite lattice. Correspondingly, we need the rigorous finite volume regularization of the topological expressions for Hall conductivity proposed in \cite{FZ2020}. To achieve this purpose we construct here the modified Buot symbols, and arrive finally at the needed definition of the finite lattice Weyl symbol. This construction is performed in three steps. At the first step we refine momentum space, i.e. consider the auxiliary momentum lattice with $2^D$ times larger number of points. The modified Buot symbol for the given lattice is defined as the Buot symbol for the refined auxiliary one. The second step is the same procedure performed with respect to coordinate space. The resulting symbol of operator is called the doubly modified Buot symbol. It appears that this symbol contains the oscillating factors in momentum space and in coordinate space. Those factors are common for {\it all} operators. Our third step, therefore, is to omit these oscillating factors. The resulting construction is the needed Weyl symbol defined on finite lattice. Fortunately, this symbol obeys all needed properties listed in Sect. \ref{SectList}. In particular, the elementary function of lattice momentum operator turns into the same function of discrete momentum, the Weyl symbol of unity is equal to real unity, etc. As a result, the limit of infinite lattice is regular, and expression for Hall conductivity approaches smoothly to the topological expression.

Using the constructed lattice Wigner - Weyl calculus we build the lattice version of Keldysh field theory for the description of the non - equilibrium processes. This theory is used to derive in a straightforward way the simple expression for the electric conductivity written in terms of the Weyl symbols of the two - point Keldysh Green function, and the Moyal products. In case of the thermal equilibrium we come to the needed expression that tends to the topological one of \cite{FZ2020} in the limit of large lattice and small temperature. This way we achieve our purpose to build the finite volume regularization of the theory described in \cite{FZ2020}.

It is worth mentioning that our expressions were derived for the case of non - interacting {\it nonhomogeneous} systems. However, we expect that, at least, for the case of thermal equilibrium at small temperature  the technique developed in \cite{ZZ2021} may be extended to the systems defined on finite lattice. As a result our expression for Hall conductivity may be, possibly, replaced by the one, in which the non - interacting Green function is replaced by the complete interacting one. The direct consideration of this possibility, however, remains out of the scope of the present paper.

Another possible extension of our results is related to rigorous description of the other non - dissipative transport phenomena. Namely, the proposed technique may be useful for the consideration of the finite volume (infrared) regularization and the consideration of the finite volume effects in Chiral Separation Effect, Chiral Torsional Effect, Spin Hall Effect, and so on. Recall that previously these effects were mainly considered for the homogeneous weak fields. Precise Wigner-Weyl calculus for the lattice models (including the models defined on finite lattices) may be useful for the rigorous description of various non - dissipative conductivities.

\acknowledgments
The author is grateful for useful discussions of various issues related to Wigner - Weyl calculus and Keldysh technique to I.Fialkovsky, M.Suleymanov, C.Zhang, M. Lewkowicz, and C.Banerjee.

\bibliography{cross-ref,wigner3,buotbibl,wigner4}

\end{document}